\documentclass[iop,onecolumn,twocolappendix,numberedappendix]{emulateapj}

\usepackage{times}
\usepackage{graphicx}
\usepackage{multirow}
\usepackage{booktabs}
\usepackage{color}
\usepackage{bm}
\usepackage{hyperref}
\hypersetup{colorlinks = true, linkcolor = blue, anchorcolor = red, 
     citecolor = blue, filecolor = red, urlcolor = blue, backref = page}

\newcommand{\Ms}{M_{\star}}
\newcommand{\Mg}{M_{\rm g}}
\newcommand{\Mbh}{M_{\rm BH}}
\newcommand{\rhos}{\rho_{\star}}
\newcommand{\rs}{r_{\star}}
\newcommand{\qs}{q_{\star}}
\newcommand{\rhog}{\rho_{\rm g}}
\newcommand{\rg}{r_{\rm g}}
\newcommand{\MR}{{\cal R}}
\newcommand{\Reff}{R_{\rm e}}
\newcommand{\phig}{\phi_{\rm g}}
\newcommand{\phiBH}{\phi_{\rm BH}}
\newcommand{\sigR}{\sigma_R}
\newcommand{\sigz}{\sigma_z}
\newcommand{\sigphi}{\sigma_{\varphi}}
\newcommand{\vphi}{v_{\varphi\star}}
\newcommand{\Lbh}{L_{\rm BH}}
\newcommand{\Ledd}{L_{\rm Edd}}
\newcommand{\Tx}{T_{\rm X}}

\begin{document}
\title{MACER improved: AGN feedback computed in rotating early-type galaxies
	at high resolution}

\author{Zhaoming Gan$^{1}$\footnotemark[$$], Luca Ciotti$^2$,  
	    Jeremiah P. Ostriker$^{3}$\footnotemark[$$], Feng Yuan$^{1}$}
\affil{$^{1}$Shanghai Astronomical Observatory, Chinese Academy of Sciences, 
	80 Nandan Road, Shanghai 200030, China}
\affil{$^{2}$Department of Physics and Astronomy, University of Bologna,
	via Piero Gobetti 93/2, 40129 Bologna, Italy}
\affil{$^{3}$Department of Astronomy, Columbia University, 
	550 W, 120th Street, New York, NY 10027, USA}


\begin{abstract}
\noindent
Based on our previous modeling of AGN feedback in isolated elliptical 
galaxies \citep{gan_active_2014} using the \texttt{MACER} 
(Massive AGN Controlled Ellipticals Resolved) code,  we extend and 
improve the model  
	to include rotation, 
	to facilitate angular momentum transfer via the Toomre instability in gaseous disks,
	to limit the star formation to regions of high density and low temperature, 
	and to improve the treatment of hot mode (low accretion rate) AGN feedback.
	The model galaxy now has an extended dark matter profile that 
	   matches with standard observations, but has a resolution of 
	   parsecs in the inner region and resolves the Bondi radius.
	
We find that the results agree reasonably well  with a panoply of observations: 
	(1) both AGN activity and star formation are primarily in 
	    central \textcolor{black}{cold gaseous} disks, 
	    are bursty and are mainly driven by the Toomre instability; 
	(2) the AGN duty cycle agrees well with the Soltan argument, 
	    i.e., the AGN spends most of its lifetime when it is in 
	    low luminosity (half of time with $L/L_{Edd}<7\times10^{-5}$), 
	    while emitting most of its energy when it is in high luminosity 
	    (half of radiant energy emitted with $L/L_{Edd}>0.06$); 
	(3) the total star formation is $\sim$ few percents of 
            the initial stellar mass, occurring in the bursts that would 
	    be associated with the observed E+A phenomenone. 
	    Most of the star formation occurs in the circumnuclear disk 
	    of a size $\le 1$ kpc, which is in agreement with
	    recent observations; 
	(4) the ISM X-ray luminosity varies within a reasonable range 
	    (median $L_{\rm X, ISM}=9.1\times10^{39}$ erg/s) 
	    in agreement with observations.
\end{abstract}


\section{Introduction}
Most of the stellar mass seen in the universe is in relatively massive 
elliptical galaxies \citep{drory_bimodal_2009} which apparently form 
at relatively high redshift as ``blue nuggets'', become quiescent 
``red nuggets'' at intermediate redshift \citep{van_dokkum_forming_2015} 
and accumulate an outer envelope 
of accreted low mass, low metallicity stars at late times 
\citep{greene_growth_2009,  naab_formation_2007}. 
Subsequent episodes of star formation and 
``E+A" phases contribute $\sim2\%$ more stars during this interval. 
The literature has been recently reviewed by \citet{somerville_physical_2015} 
and by \citet{naab_theoretical_2017}. The overall two phase evolution 
was outlined by \citet{oser_two_2010} with role of AGN (active galactic nucleus) feedback in quenching 
star formation discussed by many authors (e.g., \citealt{di_matteo_energy_2005,
springel_modelling_2005, cattaneo_agn_2007, ciotti_radiative_2007, booth_cosmological_2009, dubois_jet-regulated_2010, ostriker_momentum_2010,debuhr_growth_2011,
novak_feedback_2011, choi_radiative_2012,fabian_observational_2012, gaspari_cause_2012, dubois_agn-driven_2013, hirschmann_cosmological_2014, crain_eagle_2015, sijacki_illustris_2015, eisenreich_active_2017,hopkins_fire-2_2018, tremmel_introducing_2018, weinberger_supermassive_2018, yuan_active_2018}).

Thus for most of the observable lifetime ($z<2$) these systems have an 
evolution largely driven by internal processes.  The primary source of
 mass addition ($\sim15\%$ of the stellar mass) is from normal stellar 
evolution, while the primary energy and momentum feedback is from supernovae (SNe) Ia 
and central supermassive black holes with a sporadically important SNe II input
(important after bursts of star formation). Cosmological codes, which 
are necessary during the formation phase are ill suited to explore 
the physics of this second phase due to their low spatial resolution 
and consequent inability to model the inner several hundred parsecs 
within which  AGN feeding and feedback are determined and 
which provide the sites for starburst episodes and inner disk formation.

Over some time we have developed a high resolution mesh code to address 
this phase of galaxy evolution. The first paper in this series is 
\citet{ciotti_cooling_1997}. With over twenty papers in the series, 
we have named the steadily improving code ``\texttt{MACER}'' for Massive AGN 
Controlled Ellipticals Resolved, and this paper outlines several of 
the recent substantial code improvements now available to treat the complex 
evolution of elliptical galaxies. The code has high spatial resolution 
(parsecs in the inner regions),  standard and relatively complete 
stellar physics and chemical evolution, and implementation of AGN feedback 
that is designed to match observed BAL winds and luminous output for high accretion rates 
and a new physically modeled mode \citep{yuan_numerical_2015} 
for the low accretion (hot) outflows. 
Radiative transfer is included in a simplified, spherical Eddington 
approximation fashion. 
The purpose of this paper is to add some essential 
improvements to the physical modeling, to determine the consequences of 
these changes, and to propose observational tests which will help to
ascertain the accuracy of the improved treatments.

In brief, we find 
that in even moderately rotating normal ellipticals 
there will be periodic formation of central, cool gas disks which will 
become unstable to the classic Toomre instability 
\citep{toomre_gravitational_1964} leading to both star bursts and 
AGN feeding/feedback. 

 
\section{Model Setup}
To study black hole feeding and feedback, we need to answer two fundamental 
questions: (i) what are the mass sources? (ii) how is the mass transported 
to the galaxy center? Apparently, the mass sources could be (1) the remnant 
interstellar medium (ISM) from the galaxy formation; (2) mass accretion onto the galaxy (e.g., 
from the cosmic web); (3) stellar mass loss (e.g., stellar winds from AGBs); 
(4) recycled gas from the AGNs (e.g., BAL winds, ultra-fast outflows, hot disk winds, etc) 
and supernovae of both type I \& II. 
Provided sufficient mass 
supply, the ISM could be supported by thermal pressure and/or rotation 
(angular momentum) against the gravity, which prevents the ISM from 
being accreted too rapidly by the central supermassive black hole. 
It has been known for decades that there is nominally a strong 
``cooling flow problem'' \citep{fabian_cooling_1994}: 
the gas observed via X-rays in normal massive 
ellipticals has a radiative cooling time quite short compared to the 
Hubble time. There are energy sources available based on both the central 
supermassive black holes and normal stellar feedback that could balance
 these losses, but determining the quasi equilibrium requires a delicate 
treatment of both energy inputs and outputs since we know empirically 
that there are periodic collapses of cool gas to the center, occurring 
when energy input does not balance radiative losses, that result in AGN 
outbursts, star formation episodes and the resultant ``E+A" phenomenon
\citep{dressler_spectroscopy_1982}.

Previous papers in this series found that a slowly varying quasi equilibrium 
typically exists with gas input primarily from the AGB phase of late stellar 
evolution of the low mass stars approximately balanced by mass outflows driven 
by SNe Ia (see early work by \citealt{renzini_production_1993}). 
This quasi-equilibrium is punctuated by episodes of cooling flow 
accompanied with star formation and black hole accretion that in turn produce 
violent AGN activities and blow out gas into the CGM.

In this paper we will show how these processes are modified by the 
addition of rotation (see also \citealt{yoon_active_2018}), 
a more massive halo and a better algorithm for the 
star formation. We will find that the overall character of the evolution 
remains (quasi equilibrium interrupted by outbursts), but the geometry 
is changed significantly by the periodic formations of cool, dense, 
Toomre unstable central disks of approximately kpc size within which 
most of the regulatory activity occurs. 
In principle, star formation could be efficient enough to consume 
the ISM before it would be accreted by the supermassive black hole. 
So it is important to treat all of the processes properly in a 
self-consistent model, and it is prone to be a large-dynamical-range problem. 
It is our aim in this paper to consider all the processes above in 
one single high resolution hydrodynamical simulation, taking the galaxy 
as a "background" and tracking the fluid dynamics of the galactic medium 
during a cosmological timescale. In the rest of this section, we organize 
the context according to the physics we add in.
First of all, the evolution of the galactic medium is governed by 
the following time-dependent Eulerian hydrodynamic equations,
\begin{equation} \label{eq:massconsvr}
   \frac{\partial \rho}{\partial t} + \nabla\cdot(\rho{\bf v})
        = -\nabla\cdot\dot{\bf m}_Q  + \dot{\rho}_{\rm II}+\dot{\rho}_{\rm I}
           +\dot{\rho}_{\star} - \dot{\rho}_{\star}^{+},
\end{equation}
\begin{equation} \label{eq:momconsvr}
   \frac{\partial {\bf m}}{\partial t} + \nabla\cdot({\bf m v})
        = - \nabla p_{\rm gas} - \nabla p_{\rm rad} - \rho \nabla \phi 
           - \nabla\cdot\Pi_{\rm vis}- \nabla\cdot\Pi_Q +\dot{\bf m}_{\rm S} -\dot{\bf m}^{+}_{\star},
\end{equation}
\begin{equation} \label{eq:engconsvr}
   \frac{\partial E}{\partial t} + \nabla\cdot(E{\bf v})
        =  -p_{\rm gas} \nabla \cdot {\bf v} - \Pi_{\rm vis}:\nabla {\bf v} + H - C  + \dot{E}_{\rm Q}  
            +\dot{E}_{\rm II}+\dot{E}_{\rm I} + \dot{E}_{\rm S}-\dot{E}^{+}_{\star},
\end{equation}
\textcolor{black}{where where $\rho$, ${\bf m}$, $E$, $p_{\rm gas}$ and ${\bf v}$  are the fluid density, momentum, internal energy, thermal pressure and velocity, respectively.} $p_{\rm rad}$ is the radiation pressure of AGN irradiation due to both scattering $(\nabla p_{\rm rad})_{\rm es}$ 
and absorption 
$(\nabla p_{\rm rad})_{\rm abs}$ (\S\ref{sec:agn-feedback-model}, 
Equation \ref{eq:radiation-pressure}). 
The adiabatic index is fixed to $\gamma=5/3$.  
$\phi=\phig+\phiBH$ is the total gravitational
potential of the galaxy (stars + dark matter) $\phig$, plus that
  of the central supermassive black hole of mass $\Mbh$, $\phiBH =-G
  \Mbh/r$. The self-gravity of the gas is not taken into account (see
  \S\ref{sec:galaxy-model} for more details).

Regarding the mass sources, we treat the remnant ISM as the initial condition 
(\S\ref{sec:numerical-setup}), and treat the mass accretion onto the galaxy 
(CGM infall in our case, see \S\ref{sec:CGM-infall}) as outer boundary 
conditions, so they do not explicitly appear in the equations above. 
Our estimate for CGM infall is taken from the full cosmological zoom 
simulations \citep{choi_physics_2017, brennan_momentum-driven_2018}.
Besides, the stellar passive evolution is treated as source terms 
(\S\ref{sec:stellar-feedback}), including the mass sources and energy heating 
contributed by 
AGBs  ($\dot{\rho}_{\star} $, $\dot{E}_{\rm S} $), 
SN Ia ($\dot{\rho}_{\rm I} $, $\dot{E}_{\rm I} $), 
SN II ($\dot{\rho}_{\rm II}$, $\dot{E}_{\rm II}$) 
and also the momentum source term $\dot{\bf m}_{\rm S}$, since the stellar mass 
loss above would inherit the stream velocity (e.g., rotation, if any; 
Equation \ref{eq:stellar-rotation}) of its host stars
(please see Appendix \ref{appendix:stellar-feedback} for an outline, 
and we refer the readers to \citealt{ciotti_agn_2012} for a full description). 
We also take into account active stellar evolution, where 
$\dot{\rho}_{\star}^{+}$, $\dot{\bf m}^{+}_{\star}$, 
and $\dot{E}^{+}_{\star}$ are the mass, momentum, and energy sink terms, 
respectively, associated with star formation (\S\ref{sec:star-formation}).

It is known that SNe Ia alone are capable of heating the ISM up to the local 
Virial temperature. The hot gas can not be accreted efficiently because of 
its thermal pressure gradient and low density, so it is extremely important to evaluate 
the energy gain/loss of the ISM. In the energy equation above, $H$ and $C$ 
are the net radiative heating and cooling (under AGN irradiation), 
respectively, which include Compton heating/cooling, Bremsstrahlung cooling, 
and line heating (photoionization) /cooling (recombination). We refer 
the readers to \citet{sazonov_radiative_2005} for more details (see also 
\citealt{novak_feedback_2011,ciotti_agn_2012}), a brief highlight is also presented in 
Appendix \ref{appendix:radiation-processes} for the completeness of this paper.
We note that it falls back automatically to the case of the passive (atomic) 
cooling when the AGN luminosity is zero.

In the cases with rotation, the ISM with high angular momentum would naturally 
cool, condense and form a circumnuclear disk. The ISM on the disk can not 
be accreted without losing its angular momentum, and it will be consumed by 
star formation eventually, 
\textcolor{black}{for example, \citet{eisenreich_active_2017} found
in their SPH simulations that circumnuclear disks commonly form in the galaxy
centers, and intensive star formation occurs within those gaseous cold disks.}
  As proposed by \citet{hopkins_how_2010, hopkins_analytic_2011},
  one of the most promising mechanisms for angular momentum transfer on the 
  galactic scale is the gravitational torque due to non-axisymmetric structures 
  of the stellar population (see also \citealt{lodato_classical_2008} and references therein). 
  We assume axisymmetry in our galaxy model 
  (cf. \S\ref{sec:galaxy-model}), so we do not include the \citeauthor{hopkins_how_2010} 
  mechanism for self consistency
  (however, see \citealt{yoon_active_2018} for an alternative treatment).
However, if the gaseous disk is dense enough to become locally self gravitating, 
it is prone to be gravitational unstable (\citealt{toomre_gravitational_1964}). 
In this paper, we propose a numerical algorithm for such 
gravitational instability, in which we treat it as a diffusive process, 
where $\dot{E}_Q$, $\Pi_Q$ and $\dot{\bf m}_Q$ count for the energy dissipation, 
angular momentum transfer and mass transport, respectively, 
due to the Toomre instability (\S\ref{Q-unstable}). 
Besides, we also use the ``$\alpha$ prescription'' ($\Pi_{\rm vis}$, see \S\ref{sec:alpha-viscosity}) 
to mimic the magnetorotational instability (\citealt{balbus_instability_1998}) in the disk, 
which can also transfer angular momentum.

AGN mechanical feedback (in terms of disk winds) is treated as inner boundary 
conditions, similar to the CGM infall, i.e., injecting wind 
mass/momentum/energy at the inner boundary (Note that AGN radiation feedback 
is already included in the heating and cooling function $H-C$ as per 
earlier papers in this series; See \S\ref{sec:agn-feedback-model} 
for detailed description).

Finally, we utilize the \texttt{Athena++} code 
\citep[version 1.0.0;][]{stone_athena:_2008} to solve the hydrodynamical 
equations above. The \texttt{Athena++} code is a state-of-art, grid-based 
radiation magnetohydrodynamical code. It has flexible coordinate and 
grid options, e.g. including spherical coordinates combined with 
adaptive mesh refinement, which make it ideal 
for large dynamic-range simulations like those we've made in this paper. 
Particularly, we use spherical coordinates assuming axisymmetry while 
allowing rotation (a.k.a. 2.5-dimensional simulation). The outer boundary 
is chosen as 250 kilo-parsec to enclose the whole massive elliptical galaxy, 
the inner boundary is set to be 2.5 parsec to resolve the Bondi radius 
so that we are able to self-consistently track the black hole feeding 
processes. We use a logarithmic grid 
($\Delta r_{\rm i+1}/ \Delta r_{\rm i} = 1.1$) to divide the radial axis 
into 120 discrete cells. We subtract two small conical zones near the poles 
to avoid the well-known axial numerical singularity, the azimuthal angle 
$\theta$ is divided into 30 uniform cells and covers an azimuthal range 
from $0.05\pi$ to $0.95\pi$.
\textcolor{black}{The numerical solver for the gas dynamics is composed 
by the combination of the HLLE Riemann Solver, the PLM reconstruction 
and the second-order van Leer integrator.}


\subsection{Structure and Dynamics of the Galaxy Models} \label{sec:galaxy-model}

As is well known, the structural and dynamical properties of the galaxies are 
one of the main factors determining the gas evolution in early type galaxies (ETGs). 
In this paper we focus on the effects of large-scale ordered rotation; 
the adopted models are constructed accordingly. All the necessary steps needed 
in the construction of an axisymmetric rotating ETG (such as the determination 
of the structural parameters, the amount and distribution of DM, the recovery 
of the galaxy gravitational potential and force field, and the solution of 
the Jeans equations) can be found elsewhere (\citealp{posacki_effects_2013},
\citealt{negri_effects_2014-1}, \citealt{negri_effects_2014}, and 
in particular in \citealt{negri_x-ray_2015}, \citealt{ciotti_effect_2017} 
and \citealt{pellegrini_active_2018}), so we do not repeat here. 
In particular, in these works the galaxy models were constructed 
by solving numerically the Poisson and the two-integrals Jeans' equations.

Here instead, we use a different approach, adopting fully analytical
axisymmetric models obtained by homeoidal expansion
\citep{ciotti_simple_2005} of the two-component JJ spherical models
discussed in \citeauthor{ciotti_two-component_2018}
(\citeyear{ciotti_two-component_2018}, hereafter CZ18).  Ellipsoidal
JJ models are made by the superposition of a \citet{jaffe_simple_1983}
stellar ellipsoidal density distribution added to a DM halo so that
the {\it total} density distribution is another ellipsoidal Jaffe
models, in general with different flattening and scale length. For a
thorough discussion of the structural and dynamical properties of
these models see Ciotti \& Ziaee Lorzad (in preparation, hereafter
CZ19).  

The advantage of this approach is that it is very easy to change galaxy 
parameters (an important aspect in exploratory works) and to implement the 
relevant physical formulae in the hydrodynamical code. At the same time, 
the models agree with the main observational properties of ETGs 
(a stellar density distribution closely following the de Vaucoulers empirical 
law over a large range, a DM halo well approximated by the NFW formula, 
adjustable flattening in the stellar component and in the total mass 
distribution, a parametrized amount of ordered rotation, allowing for
the construction of galaxy models spanning the cases from tangential 
anisotropy to isotropic rotators, etc). In addition, these models  
in the case of moderate flattening (say for galaxies rounder than $E4$) 
allow for the analytical solution of the Jeans equations for the stellar
component also in presence of a central BH. Note that, being the
Jeans equations linear in the potentials, in turns this allow to update 
at each time step the values of the rotational velocity and of the
velocity dispersion tensor due to an increasing mass of the central BH.
We also notice that one could also change, as a function of time, 
the DM halo concentration, to explore the effects of halo 
contraction/expansion and BH growth on the gas flows.

The density distribution is described by an oblate
\citet{jaffe_simple_1983} model of axial ratio $\qs$, total mass $\Ms$
and scale-length $\rs$:
\begin{equation} \label{eq:JJ2}
\rhos = \frac{\Ms}{4\pi \rs^3}\frac{1}{\qs m^2(1+m)^2},\quad
m^2 = s^2\left(\sin^2\theta+\frac{\cos^2\theta}{\qs^2}\right), 
\end{equation}
where $s\equiv r/\rs$ and $(r,\theta,\varphi)$ are the standard
spherical coordinates. In spherical models ($\qs=1$), $\Reff\simeq
0.75~\rs$, where $\Reff$ is the effective radius of the galaxy; in the
edge-on projection of oblate models, $\Reff\simeq 0.75\sqrt{\qs}~\rs$.

In JJ models we then assign the total galaxy density (stars + dark
matter) $\rhog$, so that the resulting DM halo is given by the
difference $\rhog -\rhos$.  Here for simplicity we restrict to the
case of a spherical total density $\rhog$, given by a spherical Jaffe
profile of total mass $\Mg =\MR\Ms$ and scale-length $\rg=\xi\rs$, so
that
\begin{equation} \label{eq:total_density}
\rhog = \frac{\Ms}{4\pi \rs^3}\frac{\MR\xi}{s^2(\xi+s)^2}.
\end{equation}
The positivity request of $\rhog$ imposes constraints on the values of 
$\MR$ and $\xi$, and in CZ18 it is shown that in the {\it minimum halo models}
(as those here used), the DM profile is described very well by the NFW profile.  

The total gravitational potential of the galaxy plus the central MBH of mass 
$\Mbh=\mu\Ms$ is then given by 
\begin{equation} \label{eq:total_grav_pot}
\phi=\phig +\phiBH = {G\Ms\MR\over \rs\xi} \ln\left(\frac{s}{s+\xi}\right) -
{G\Ms\mu\over \rs s}.
\end{equation}
The circular velocity in the equatorial plane is given 
\begin{equation} 
\frac{v_c^2}{r} = \frac{d\phi}{dr}.
\label{eq:circular_velocity}
\end{equation}
In CZ19 the Jeans equations are solved, and it shows that the radial 
and vertical velocity dispersions, $\sigR=\sigz$, can be written as
\begin{equation}  \label{eq:stellar-velocity-dispersion-part1}
	\rhos\times\sigR^2 = {G\Ms^2\over 4\pi \rs^4}\times
\left\{
\mu\left[A(s)+\eta B(s) + \eta C(s)s^2\sin^2\theta\right]+
{\MR\over\xi}\left[D(\xi,s)+\eta E(\xi,s) + \eta
  F(\xi,s)s^2\sin^2\theta\right]
\right\},
\end{equation}
where $\eta=1-\qs$ and the radial function are simple analytical
functions, and the separate contributions of the central MBH and of
the galaxy potential to the velocity dispersion is apparent. For
spherical JJ models, the central projected velocity dispersion of
stars, due to the galaxy contribution only, is given by $\sigma_{\rm
  p}^2(0) = G\Ms\MR/(2\rs\xi)$, and this is a very good approximation
also for the ellipsoidal models for low flattening.

\begin{figure}[htb]
\centering
\includegraphics[width=0.6\textwidth]{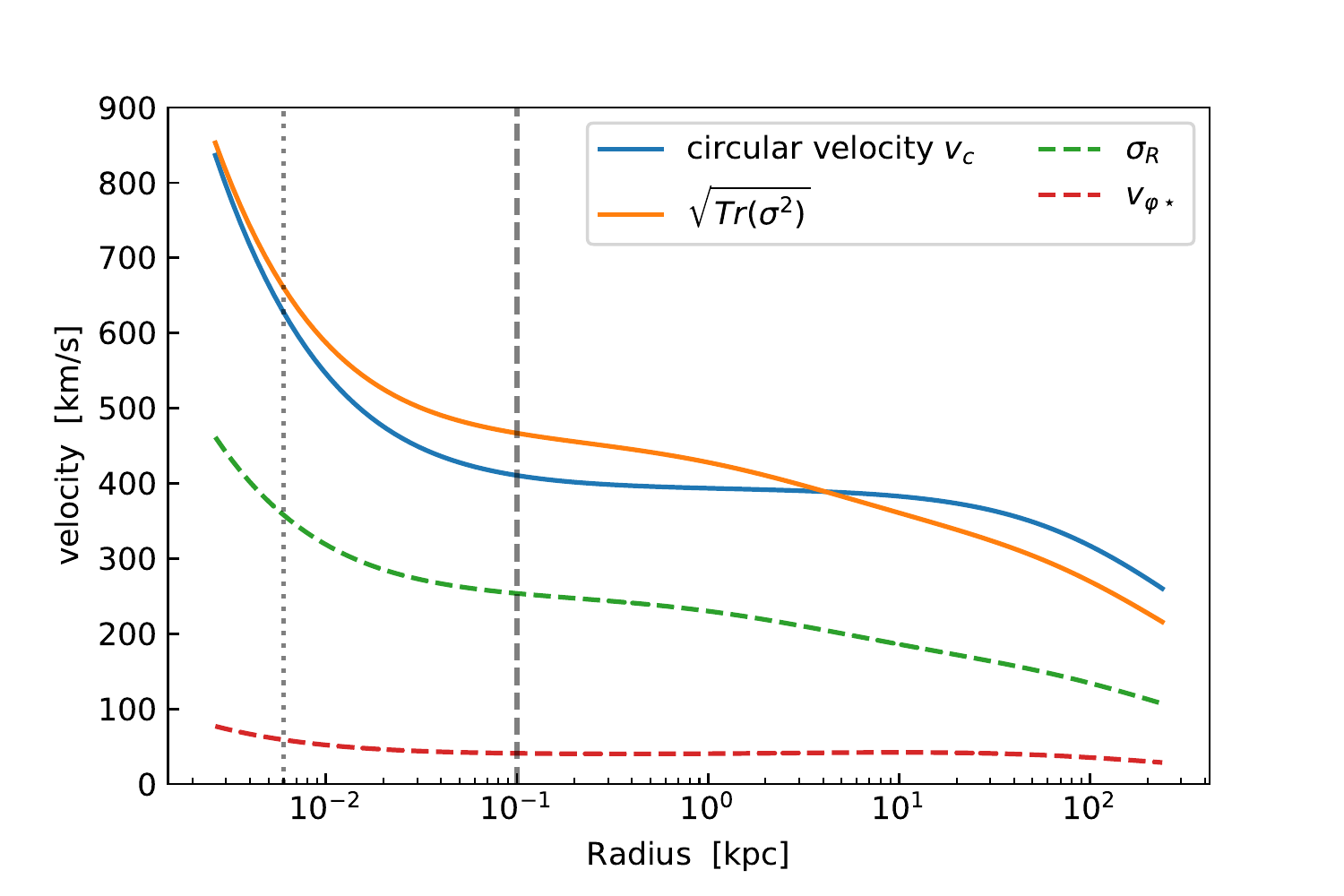}
\caption{Characteristic velocities (on the equatorial plane, i.e., $\theta=\pi/2$) of the model galaxy: 
(1) circular velocity ($v_c$, blue line; which characterizes the total gravitational potential); 
(2) square root of the trace of stellar velocity dispersion ($\sqrt{{\rm Tr}(\sigma^2)}$, orange line; 
	which determines the stellar thermalization); 
(3) radial stellar velocity dispersion ($\sigma_R$, green dashed line); and 
(4) the ordered rotation velocity of stars ($\vphi$, red dashed line; 
	which determines the specific angular momentum of the stellar mass loss). 
The galaxy model parameters adopted are $\Ms = 3.35\times10^{11} M_{\odot}$, $\rs = 9.3$ kpc, 
$\MR =20$, $\xi=20$, $\mu=10^{-3}$, $\eta=0.2$ and $k=0.25$.
The vertical dotted line shows approximately the Bondi radius ($\sim6$ pc, assuming a typical ISM temperature of $10^7$ K). 
The vertical dashed line shows approximately the radius of influence of the black hole ($\sim100$ pc). 
Inner boundary of our simulations is 2.5 pc. }
\label{fig:velocity-dispersion}
\end{figure}

As is well known, the two-integrals Jeans equations are degenerate, i.e., 
they just provide the total (ordered plus velocity dispersion) kinetic energy 
in the azimuthal direction, via the quantity 
$\Delta_\star =\vphi^2 + \sigphi^2 - \sigR^2$, where $\vphi$ is the ordered 
(i.e., streaming) velocity field of stars. To break the degeneracy we adopt 
the usual Satoh (1980) decomposition (even though more complicate 
decompositions could be used, e.g., see \citealt{ciotti_energetics_1996, negri_effects_2014-1}), with 
\begin{equation} \label{eq:stellar-rotation}
\vphi^2 = k^2\Delta_\star. 
\end{equation}
In CZ19 it is shown that for the present models
\begin{equation}  \label{eq:stellar-velocity-dispersion-part2}
	\rhos\times\Delta_\star = {G\Ms^2\eta\,s^2\sin^2\theta\over 2\pi \rs^4}\times
\left[\mu\,C(s)+{\MR\,F(\xi,s)\over\xi}\right];
\end{equation}
the explicit form of the radial functions $A-F$ is given in CZ19.
Finally, we can obtain the trace of the velocity as
\begin{equation} \label{eq:stellar-velocity-dissipation}
{\rm Tr}(\sigma^2) = 3 \sigR^2 + (1-k^2)\Delta_\star . 
\end{equation}
With the ordered and dispersive velocity field, we are able to evaluate 
the specific angular momentum and also the stellar thermalization 
(see Appendix \ref{appendix:stellar-feedback} for details).
In the fiducial setup, we study a massive rotating elliptical galaxy
with total stellar mass $\Ms = 3.35\times10^{11} M_{\odot}$ (assuming
a mass-light ratio of 5.8 in the solar unit, scale radius $\rs = 9.3$
kpc, $\MR =20$, $\xi=20$, $\eta=0.2$ and $k=0.25$). The resulting
estimate for the central projected velocity dispersion of stars
(without the BH contribution) is therefore $\simeq 280$ km/s, placing
the galaxy model on the observed scaling laws of ETGs. The initial
black hole mass $\Mbh$ is set to $3.35\times10^8 M_{\odot}$ (i.e.,
$\mu=10^{-3}$; \citealt{magorrian_demography_1998,kormendy_coevolution_2013}). 
In Figure \ref{fig:velocity-dispersion}, we plot the
velocity profiles (on the equatorial plane, i.e., $\theta=\pi/2$)
derived from the total gravitational potential.

\subsection{Stellar Feedback}\label{sec:stellar-feedback}
The model galaxy is assumed to be 2 Gyr old at the beginning of our 
simulations, and we track the ISM dynamics for a time span of 12 Gyr. 
During such a period, the total stellar mass loss is $\sim10\%$ of 
the initial stellar mass, which alone is far more than enough to fuel 
the central supermassive black hole.  We consider both Type Ia supernova 
feedback from the old stellar population and Type II supernova feedback 
from the newly formed stars during the simulations, as they are important 
energy sources to heat the ISM. Of course, we need an initial mass function 
to evaluate the overall stellar evolution, and we adopt the Salpeter profile.

Numerically, we treat the stellar feedback as source terms, i.e., inject 
the mass in-situ where it is produced, and assume it inherits the velocity 
of its host stars \citep{ho_origin_2009}. Apparently, the kinetic energy due to the velocity 
dispersion would be quickly thermalized (a.k.a., stellar thermalization), 
and the ISM could be heated to a temperature nearly equal to the local Virial 
temperature. The angular momentum (ordered rotating velocity field) 
should be conserved (see Appendix \ref{appendix:stellar-feedback} for details).

\subsection{CGM Infall}\label{sec:CGM-infall}
\begin{figure}[htb]
\centering
\includegraphics[width=0.6\textwidth]{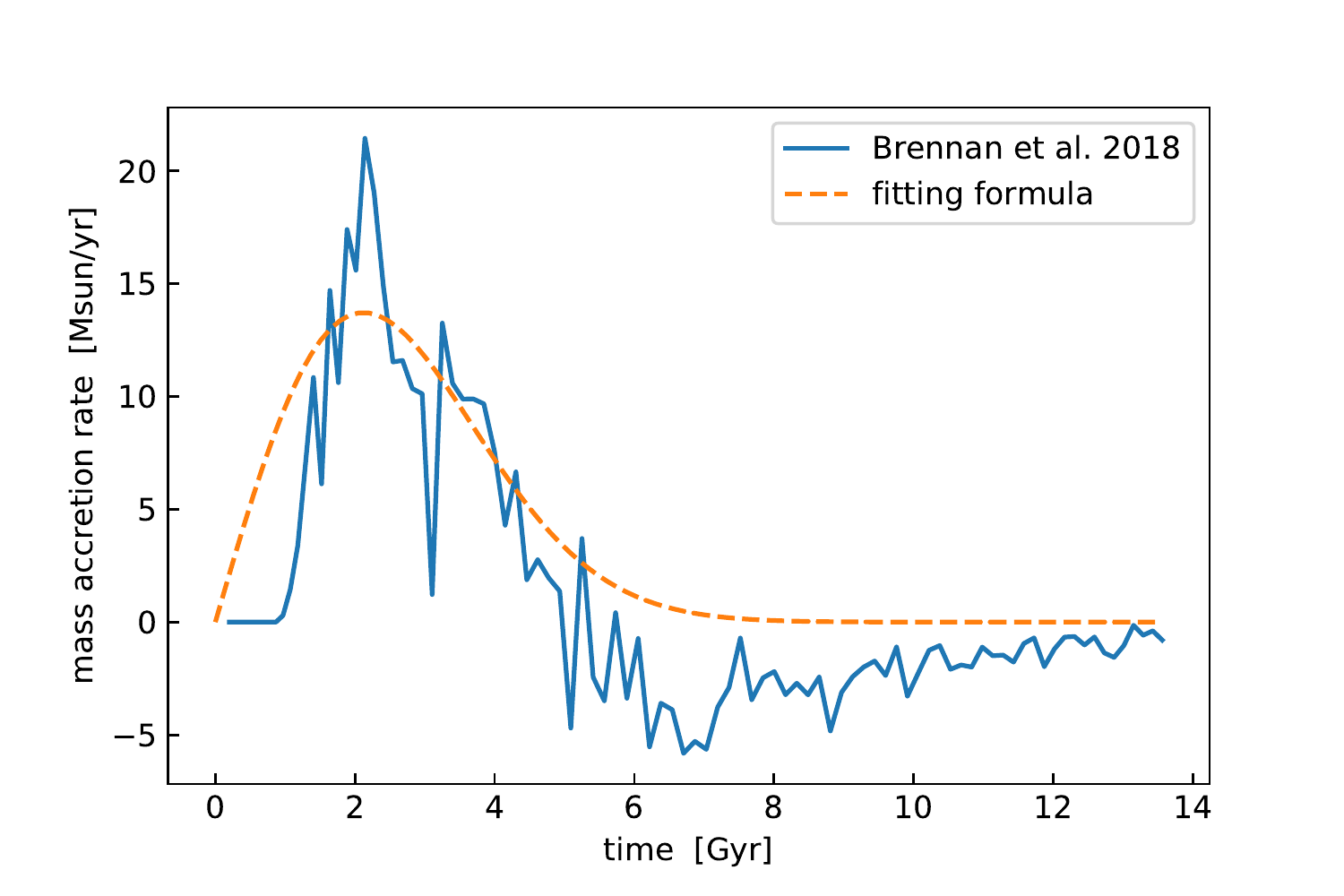}
\caption{Mass accretion rate of the CGM infalling onto the galaxy outskirt.  
The net mass accretion profile (blue solid line) is adopted from a large cosmological 
zoom simulation by \citet{brennan_momentum-driven_2018}. 
In our simulations, we use the fitting formula (Equation \ref{eq:CGM_infall},
orange dashed line) with total accreted mass 8.3\% of the initial stellar mass.}
\label{fig:CGM_infall}
\end{figure}

It is known that the mass accretion from the cosmic web can be very 
significant when compared to the ISM content remaining in the galaxies. 
We adopt the gas accretion profile onto the elliptical galaxies in a cosmological zoom-in simulations from \citet{brennan_momentum-driven_2018}. We take the mean accretion rate of 30 central elliptical galaxies, with mean stellar mass of  $2\times10^{11} M\odot$ 
\textcolor{black}{at z=0} \citep{choi_physics_2017}.
In Figure \ref{fig:CGM_infall} 
we show the normalized mass accretion rate versus time. In our numerical setup, 
we fit the profile as follows,
\begin{equation}     \label{eq:CGM_infall}
  \dot{M}_{\rm CGM} = \frac{M_{\rm acc}}{t_0/2 \cdot [1-e^{-(\Delta t/t_0)^2}]} 
  				\cdot  (t/t_0) \cdot e^{-(t/t_0)^2}
\end{equation}
where $t_0=3$ Gyr, and ${M}_{\rm acc}$ is the total mass accreted 
during the time span of $\Delta t=12$ Gyr. 
\textcolor{black}{We scale ${M}_{\rm acc}$ according to the total stellar mass of the modeling galaxy (see below for details), and we can see that the CGM infall occurs mainly in the early epoch when $t<6$ Gyr.}

Numerically, we inject the CGM at the outer boundary of the computational 
domain, which can be considered as a boundary condition. 
\textcolor{black}{The injected CGM is assumed to be nearly free-fall 
(with a constant $v_r=-(2/3) v_c$ and zero rotation velocity over the boundary 
at $r=250$ kpc). 
Its sound speed is assumed to be $c_s^2= (5/9)v_c^2$. 
The parameters above are chosen to make sure that the infalling gas 
is bound to the galaxy gravity.
The CGM inflow flux is weighted by $\sin^2(\theta)$, 
i.e., most of the CGM is injected near the equatorial plane. 
The total mass infall is scaled according to the stellar mass $\Ms$, and it is}
taken to be $M_{\rm acc} = 2.8\times10^{10} M_\odot(=8.3\%\Ms)$, 
which is comparable to the total stellar mass loss,
and approximately two times the mass of the initial ISM (\S\ref{sec:numerical-setup}).

\subsection{Circumnuclear Disk and Toomre Instability} \label{Q-unstable}
\begin{figure*}[htb]
\centering
\includegraphics[width=0.375\textwidth]{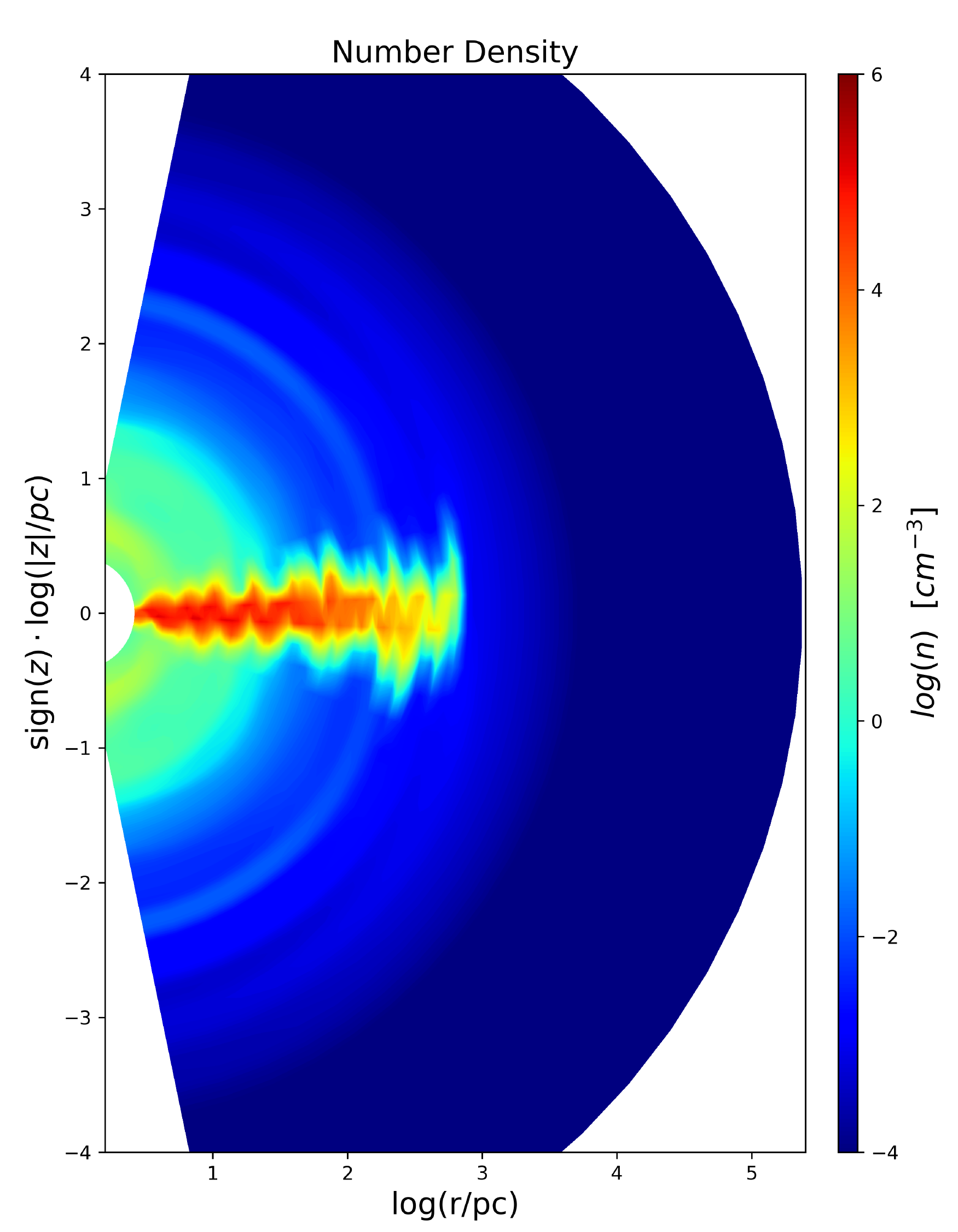}
\includegraphics[width=0.375\textwidth]{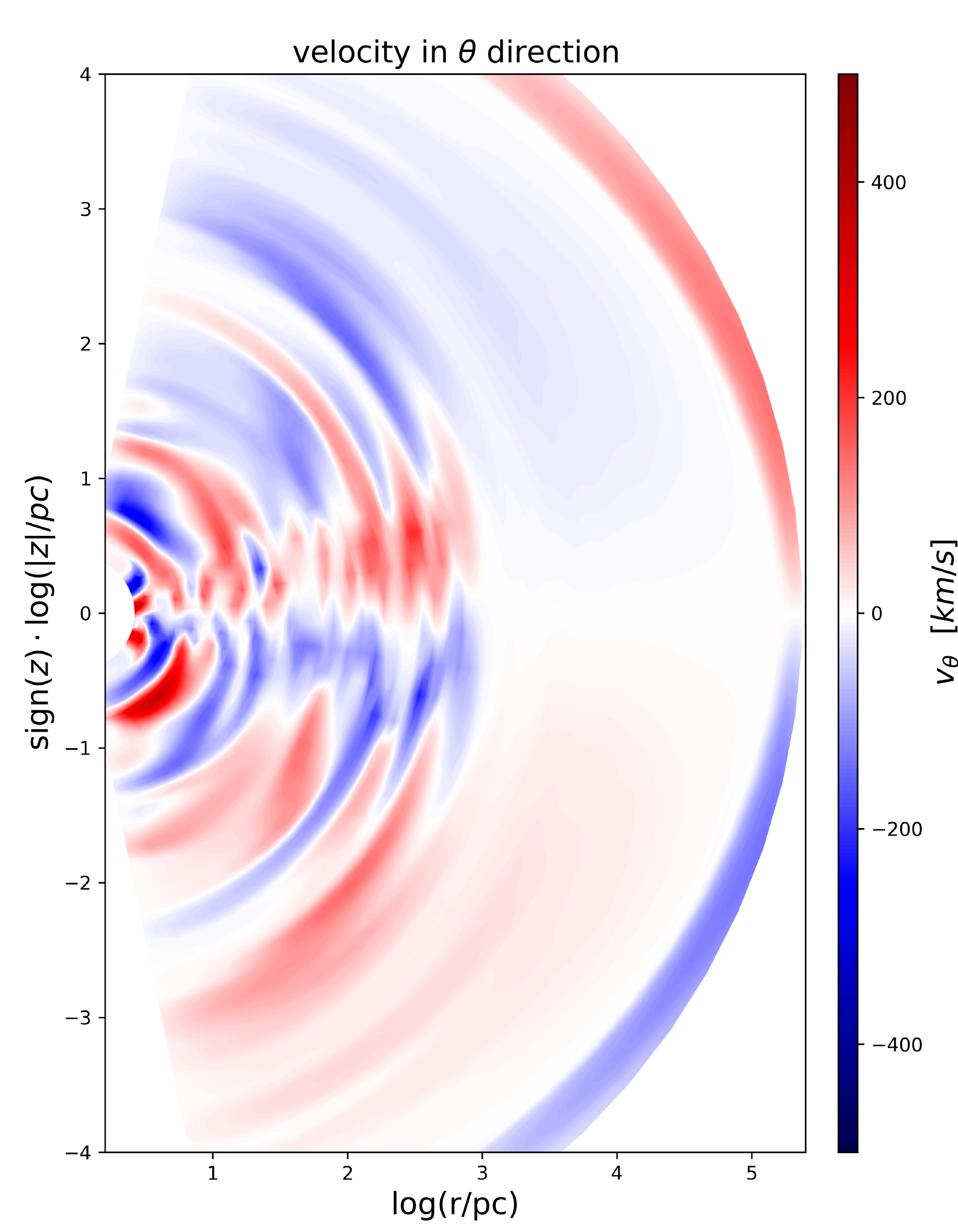}
\caption{Circumnuclear disk formed in the fiducial run (at the end of the simulation). Left panel: pseudo-color map of the ISM number density (which is wiggling); Right panel:  pseudo-color map of the azimuthal velocity $v_\theta$ (the ISM near the equatorial plane is collapsing onto the disk). The cold dense inner disk has a radius of several hundred parsecs. Note the logarithmic radial scale.}
\label{fig:cold-disk}
\end{figure*}

As ISM accumulates in the galaxy either from the stellar mass loss and/or 
the CGM infall, it would be subject to a cooling flow. When the ISM cools down 
via radiation cooling, a cold disk forms because of the angular momentum 
barrier (as shown in Figure \ref{fig:cold-disk}). 
	\textcolor{black}
	{SAURON and ATLAS have reported on the observations of cold gaseous discs 
	 in over half of the observed ellipticals 
	 (\citealt{sarzi_sauron_2006,davis_atlas3d_2011};
	 see also \citealt{boizelle_alma_2017} for ALMA observations).
	 } 
It is well known that such a cold dense disk can become gravitationally unstable if its surface 
density is greater than some critical value (a.k.a., the Toomre instability). 
The Toomre criterion considers that the disk is unstable \citep{toomre_gravitational_1964} 
when
\begin{equation} \label{eq:Toomre-Q}
 Q \equiv \frac{c_s \kappa}{\pi G \Sigma} < 1,
\end{equation} 
where $\Sigma$ is the surface density of the disk, and $\kappa$ is 
the local epicyclic frequency, 
\begin{equation}
 \kappa^2 \equiv \frac{2 \Omega}{r}\frac{d(r^2 \Omega)}{dr}, 
 \quad\quad {\rm where} \quad \Omega={v_c \over r}.
\end{equation}

As a consequence of the gravitational instability,  spiral waves will be 
developed in the cold disk, which are capable of  {\it transferring angular 
momentum outward}  by virtue of the non-axisymmetric gravitational torque 
and at the same time  {\it transferring mass inward}. The typical timescale 
is around the local orbital time. 
\textcolor{black}{However, limited by our two-dimensional settings, 
we take in account the effect of the Toomre instability by proposing 
a semi-analytical algorithm, while leave solving the self-gravity 
of the gas in a full three-dimensional simulation to our future work.}
To mimic such a process of angular momentum 
transfer, we propose a numerical algorithm as follows, 
\begin{enumerate}
\item We sample the disk density vertically (as the cold disk is 
	geometrically thin, we sample along the $\theta$ direction 
	for simplicity). Then we could evaluate the disk surface density 
	and finally determine the Toomre Q parameter of each disk ring. 
\item When a disk ring becomes unstable ($Q<1$), we move the ring 
	inward at a rate below and calculate the mass flux 
	($\dot{\bf m}_Q$) accordingly.
	\begin{equation} \label{eq:Delta-Q}
 	\frac{dr}{dt} = \frac{\Delta Q\cdot r}{\pi r/v_c}, 
	\quad\quad {\rm where} \quad \Delta Q = {\rm max}(1-Q, 0).
	\end{equation}
\item We assume the gas inherits the temperature and velocity of 
	the inner adjacent ring. To conserve angular momentum, 
	we dispose of the excessive angular momentum ($\Pi_Q$) in
	the outer adjacent ring. To conserve energy, we dissipate 
	the thermal energy gain $\dot{E}_Q$  into the inner, 
	local and outer rings according to a partition of 1/4, 1/2 
	and 1/4, respectively. 
	So, locally the angular momentum and mass transfer rates 
	are proportional to $\Delta Q$, 
	and mass/angular momentum/energy are all conserved.
\end{enumerate}

\subsection{Viscosity in the Circumnuclear Disk} \label{sec:alpha-viscosity}
As in black hole accretion disks, there should also be an effective viscosity in the circumnuclear disk due to the magnetorotational instability (MRI, \citealt{balbus_instability_1998}). To count in the magnetic process in our hydrodynamic simulations,  we use the ``$\alpha$ prescription" \citep{shakura_black_1973} to mimic the MRI effects in transferring angular momentum (e.g. \citealt{takasao_three-dimensional_2018, zhu_global_2018}).  
The viscosity coefficient reads,
\begin{equation} \label{eq:alpha-vis-coeff}
 \nu = \alpha \cdot c_s^2 /\Omega,
\end{equation} 
where $c_s$ is the sound speed as usual. To confine the viscous effects to within the circumnuclear disk, we propose a profile of the dimensionless viscosity parameter $\alpha$ as below,
\begin{equation} \label{eq:alpha-profile}
 \alpha = \alpha_0 \cdot \exp\left[-3\left(\frac{v_\varphi-v_c}{v_\varphi}\right)^2\right],
\end{equation}
where $(v_\varphi,v_c)$ are the actual and circular rotational velocities of the gas, respectively. 
We adopt $\alpha_0=0.03$ as indicated by magnetohydrodynamical simulations \citep{zhu_global_2018}, i.e., $\alpha$ is constant within the rotation-supported disk, while it decays rapidly if it is off the disk.
Following \citet{stone_hydrodynamical_1999}, we assume the azimuthal components of the viscous shear tensor $\Pi_{\rm vis}$ are non-zero, i.e., 
\begin{eqnarray}
\Pi_{\rm vis,r\varphi} &=& - \rho\nu \cdot \frac{\partial}{\partial r}\left(\frac{v_\varphi}{r}\right), \\
\Pi_{\rm vis,\theta\varphi} &=& - \rho\nu \cdot \frac{\sin \theta}{r}\frac{\partial}{\partial \theta}
                                                 \left(\frac{v_\varphi}{\sin \theta}\right).
\end{eqnarray}

From the equations above, we can see that the ratio between the viscosity timescale $\tau_{\rm vis}$ and the local orbital timescale $\tau_{\rm rot}$ scales approximately as 
$\tau_{\rm vis}/\tau_{\rm rot} \propto  (1/\alpha)(v_\varphi/c_s)^2$.  As the temperature of the cold circumnuclear disk is far below the local Virial temperature, i.e., $c_s \ll v_c \sim v_\varphi$, the viscosity timescale is usually much longer than $\tau_{\rm rot}$. Recalling that the timescale of the Toomre instability is comparable to $\tau_{\rm rot}$ (cf. Equation \ref{eq:Delta-Q}), so the angular momentum transfer is usually dominated by the Toomre instability in our model setup if parts of the disk are dense enough to become gravitationally unstable.  

The angular momentum transfer mechanisms above make it possible for the gas 
in the circumnuclear disks to be accreted by the central supermassive 
black hole and to trigger AGN activities.

\subsection{Star Formation} \label{sec:star-formation}
Another consequence of gravitational instability is star formation, i.e., 
when the ISM cools down and becomes dense enough, it is subject to the 
Jeans instability and would trigger star formation. However, it is also known 
that rotation would somehow stabilize the flow and suppress star formation. 
Similarly, we propose the star formation rate in the disk as follows when it is 
Toomre unstable,
\begin{equation}
 \dot{\rho  }_{\star,Q}^{+} = \eta_{\rm SF,Q} \cdot \Delta Q \cdot \rho \cdot \Omega 
                 \textcolor{black}{
                 = \eta_{\rm SF,Q} \cdot \Delta Q \cdot \rho \cdot  \sqrt{4\pi G \bar{\rho}/3}.
                 }
  \label{eq:rhodot_sf_Q}
\end{equation}
\textcolor{black}{where $\bar{\rho}$ is the mean density interior to the radius.}
We assume $\eta_{\rm SF,Q}=0.1$. When turning gas into stars, we simply 
subtract the amount of gas with the in-situ fluid velocity and temperature, 
and replace it with stars without changing the specific energy 
and momentum per unit mass,
\begin{equation}
 \dot{\bf m}_{\star,Q}^{+} = \frac{\dot{\rho  }_{\star,Q}^{+} }{\rho}\cdot {\bf m}, \quad\quad
 \dot{E     }_{\star,Q}^{+} = \frac{\dot{\rho  }_{\star,Q}^{+} }{\rho}\cdot E.
\end{equation}

Throughout the galaxy we also evaluate the star formation due to the local Jeans instability 
(though it is suppressed by rotation). More specifically \citep{ciotti_agn_2012},
\begin{equation}
 \dot{\rho  }_{\star,C}^{+} = \frac{\eta_{\rm SF,C} \rho     }{\tau_{\rm SF,C}}, \quad\quad
 \dot{\bf m}_{\star,C}^{+} = \frac{\eta_{\rm SF,C} {\bf m} }{\tau_{\rm SF,C}}, \quad\quad
 \dot{E     }_{\star,C}^{+} = \frac{\eta_{\rm SF,C} E          }{\tau_{\rm SF,C}}.
  \label{eq:rhodot_sf_C}
\end{equation}
We set a low star formation efficiency $\eta_{\rm SF,C}=0.01$, 
and $\tau_{\rm SF,C} = \max(\tau_{\rm cool}, \tau_{\rm dyn})$, where
\begin{equation}
  \tau_{\rm cool} = {E\over C}, \quad
  \tau_{\rm dyn} = \min(\tau_{\rm Jeans}, \tau_{\rm rot}),\quad
  \label{eq:tau_sf1}
\end{equation}
In addition, we do not allow star formation when the gas density is lower 
than $10^5~{\rm atom}/{\rm cm}^{-3}$ nor the gas temperature is higher 
than $4\times10^4$ K. Although we do not include the formation of 
molecular gas, our very high threshold for star formation, which is possible, 
given our high resolution, is comparable to the density in star forming 
molecular clouds.

The total star formation rate is the sum of Equation \ref{eq:rhodot_sf_Q} 
and \ref{eq:rhodot_sf_C},
\begin{equation}
 \dot{\rho  }_{\star}^{+} = \dot{\rho  }_{\star,Q}^{+} + \dot{\rho  }_{\star,C}^{+}, \quad\quad
 \dot{\bf m}_{\star}^{+} = \dot{\bf m}_{\star,Q}^{+} + \dot{\bf m}_{\star,C}^{+}, \quad\quad
 \dot{E     }_{\star}^{+} = \dot{E     }_{\star,Q}^{+} + \dot{E     }_{\star,C}^{+}.
  \label{eq:star-formation-rate}
\end{equation}

The cold circumnuclear disk is the fuel reservoir for both star formation 
and black hole accretion, and there is a tough competition between 
these two processes. For example, if star formation is very efficient, 
most (if not all) of the cold gas will be consumed before it could 
be accreted by the supermassive black hole, so the AGN activities will be 
significantly suppressed. Vice versa, if angular moment transfer 
is very efficient, strong AGN feedback will be triggered, which in turn 
will suppress star formation further. However, both star formation and 
angular momentum transfer are related to the same physics, it means that 
they are of similar timescales, i.e., the timescale of the Toomre instability. 
It would be very interesting to study the balance between the two 
important processes.

\subsection{AGN Feedback} \label{sec:agn-feedback-model}
Our model for AGN feedback is founded on the concept that we should model 
the physics as closely as possible on the observed electromagnetic and 
wind outputs. AGN radiation feedback interplays with the system by affecting
 the radiation source terms $H-C$. AGN wind feedback is implemented 
by injecting wind mass, momentum, and energy via the inner boundary directly, 
so it does not implicitly appear in the equations above. In the numerical setup,
radiation feedback is determined by the AGN luminosity $L_{\rm BH}$ and 
its Compton (radiation) temperature $T_X$, while wind feedback is 
characterized by the wind mass loading rate $\dot{M}_w$ and 
its velocity $v_w$, which are determined by the AGN sub-grid model below 
(we refer the readers to \citealt{ostriker_momentum_2010} 
and \citealt{yuan_active_2018} for more details. 
\textcolor{black}{Note that we do not yet include the feedback effect of collimated jets. We leave it to our future work as the underlying physics of jet feedback on the galaxy scale is still an important open question (see, e.g., \citealt{nesvadba_extreme_2006,  nesvadba_compact_2007, nesvadba_sinfoni_2017, salome_cold_2006, guo_feedback_2008, tortora_agn_2009, wagner_driving_2012, hitomi_collaboration_quiescent_2016, yang_how_2016, Zhuravleva_nature_2016, fabian_sound_2017}).}
For completeness of this paper, we briefly introduce the model setup 
and highlight the improvement we have made).

By solving the time-dependent Eulerian equations 
([\ref{eq:massconsvr}]-[\ref{eq:engconsvr}]), we can track the mass inflow 
across the inner boundary, which would fall into the galaxy center 
(after a timescale $\tau_{\rm infall} \sim 3\times10^3$ year --- 
from the inner boundary to the black hole accretion disk) and eventually 
form a black hole accretion disk (assuming a disk size of $2000 G \Mbh/c^2$, 
which gives an accretion timescale $\sim800$ year --- 
from the disk to the black hole horizon).  After considering the time lags 
above, we obtain the disk accretion rate $\dot{M}_{\rm disk}$, 
based on which ($\dot{M}_{\rm disk} = \dot{M}_{\rm BH} + \dot{M}_{w}$) 
we finally evaluate the black hole accretion rate $\dot{M}_{\rm BH}$ 
(so $L_{\rm BH}$, $T_X$), and also the nuclear wind properties 
($\dot{M}_{w}$, \textcolor{black}{$v_{w}$}) according to our knowledge of 
black hole accretion theory and observed outflows.


We use the two-mode black hole accretion scenario \citep{yuan_active_2018} 
to achieve closure of the sub-grid model, which gives the relation 
between $\dot{M}_{\rm BH}$ and $\dot{M}_{w}$: 
(1) when the mass supply is sufficient, it will result in a relatively large value for 
$\dot{M}_{\rm disk}$, i.e., the density within the accretion disk could be 
high enough to make it radiatively efficient \citep{shakura_black_1973}, 
then, it should be in the cold (quasar) mode; 
(2) otherwise, it is in the hot mode, i.e., the radiatively inefficient mode 
but with strong wind (i.e., mass outflow; see, e.g.,
\citealt{
stone_hydrodynamical_1999,
yuan_numerical_2012,
narayan_grmhd_2012,
li_rotating_2013,
yuan_numerical_2015}). 
In the simulations, we switch the AGN sub-grid model between the cold and 
hot modes according to a critical disk accretion rate 
$\dot{M}_{\rm disk, crit} = 0.02 \dot{M}_{\rm Edd}$ 
(where $\dot{M}_{\rm Edd} \equiv L_{\rm Edd}/0.1c^2$, and 
$L_{\rm Edd}$ is the Eddington luminosity; see 
\citealt{yuan_hot_2014} and references therein), i.e., 
when $\dot{M}_{\rm disk}\ge\dot{M}_{\rm disk, crit}$ we set it to 
the cold mode, otherwise we switch it to the hot mode.

\begin{figure*}[htb]
\centering
\includegraphics[width=0.6\textwidth]{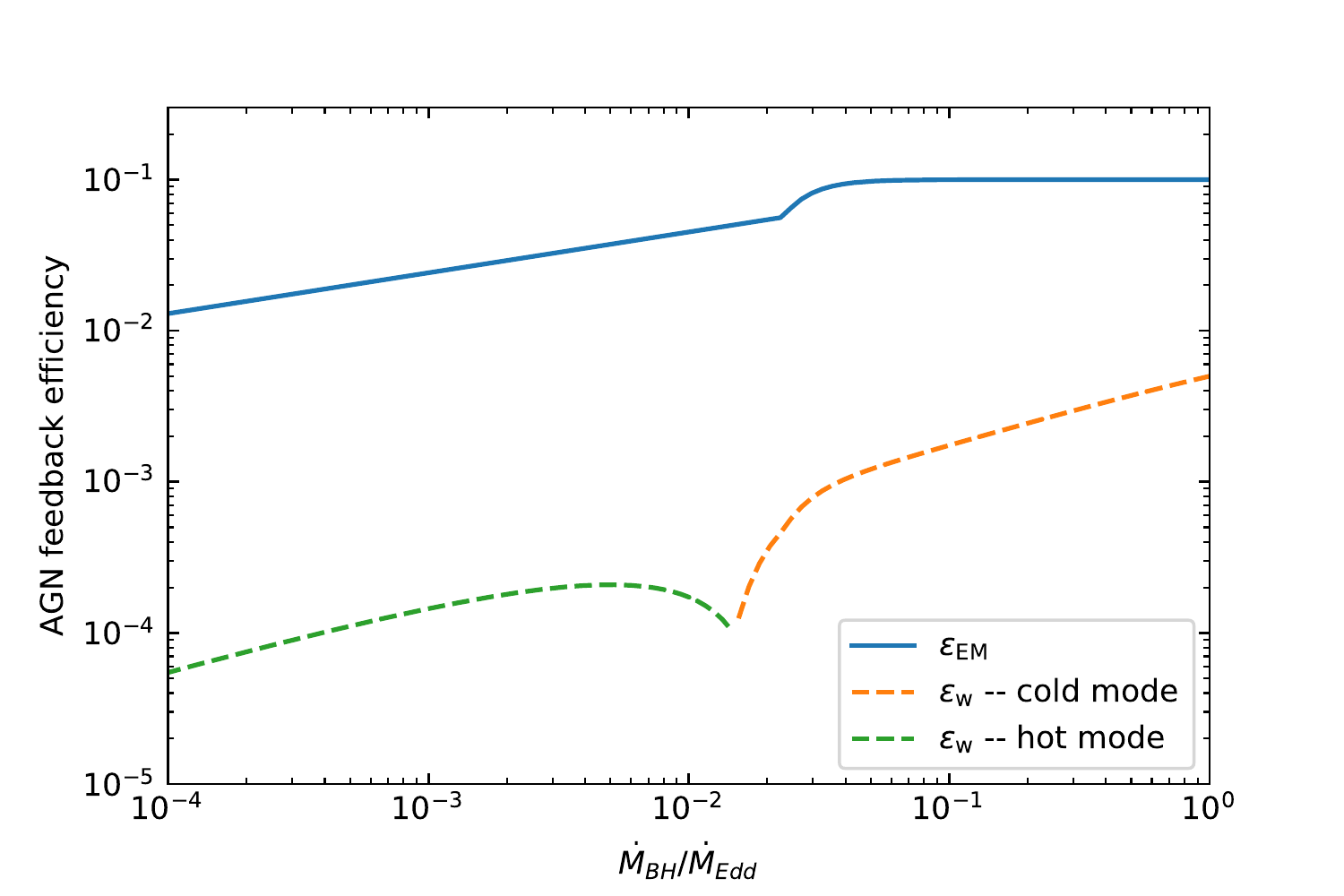}
\caption{The radiative efficiency (blue sold line) and wind efficiency (dashed lines) adopted in our two-mode AGN feedback (sub-grid) model. The mode transitions at $\dot{M}_{\rm BH} \simeq 0.02 \dot{M}_{\rm Edd}$ 
(see, e.g., \citealt{yuan_hot_2014}). Both curves are designed to approximately match observed feedback efficiencies.}
\label{fig:eps-rad}
\end{figure*}

Following the notations in \citet{ostriker_momentum_2010}, the AGN wind 
energy and momentum can be written as functions of its mass loading rate 
$\dot{M}_{w}$ and the wind velocity $v_w$,
\begin{equation}
  \dot{E}_{w} = \frac{1}{2} \dot{M}_{w} v_w^2 \equiv \epsilon_w \dot{M}_{\rm BH}  c^2  ,
  \quad\quad  \dot{P}_{w} = \dot{M}_{w} v_w.
\end{equation}

In the cold (high accretion rate) mode, we assume a constant wind velocity of
$10^4$ km/s, while allowing the wind feedback efficiency $\epsilon_w$ 
to vary as a function of the dimensionless AGN luminosity 
$l$ with a characteristic value of 
$ \epsilon_w^M = 5\times10^{-3}$ \citep[as in][]{ostriker_momentum_2010}, 
i.e., 
\begin{equation}
	\dot{M}_{w} = {2 \epsilon_w \dot{M}_{\rm BH} c^2\over v_w^2}, 
	\quad\quad v_w = 10^4 ~~{\rm km/s},
\end{equation}
where
\begin{equation}
    \epsilon_{w} =  \epsilon_w^M \sqrt{\frac{5}{4}\cdot\frac{ l}{1+l/4}
         \cdot e^{-\left({\dot{M}_{\rm disk,crit}}/{\dot{M}_{\rm BH}}\right)^4}},
    \quad\quad l\equiv {\Lbh\over \Ledd}.
\end{equation}
This is estimated to roughly match observations of BAL outflows
\citep{arav_radiative_1994,arav_measuring_2008}.

In the hot (low accretion rate) mode, the wind mass loading rate and 
the wind velocity are determined by the ``truncation" radius $r_{tr}$ 
(i.e., the outer boundary of the hot accretion disk; 
\citealt{yuan_numerical_2015}). 
\begin{equation}
  \dot{M}_{w}  = \dot{M}_{\rm disk}\cdot\left( 1- \sqrt{{3r_s\over r_{\rm tr}}}\right), \quad
   \textcolor{black}{v_{w}  = 0.1\sqrt{{G \Mbh\over r_{\rm tr}}}}, 
\end{equation}
where $r_s = 2 G \Mbh/c^2$, and 
\begin{equation}
    \textcolor{black}
    {r_{\rm tr}  = 3 r_s \left(\frac{\dot{M}_{\rm disk,crit}}{\dot{M}_{\rm disk}}\right)^2.}
\end{equation}

By solving the AGN sub-grid model above, we could get $\dot{M}_w$, 
$\dot{P}_w$, $\dot{E}_w$ and $\dot{M}_{\rm BH}$. Then, we translate 
$\dot{M}_{\rm BH}$ to the AGN luminosity $\Lbh$ by assuming 
the radiation efficiency $\epsilon_{\rm EM}$ as follows 
(\citealt{xie_radiative_2012}; see Figure \ref{fig:eps-rad}),
\begin{equation}
  \Lbh = \epsilon_{\rm EM} \dot{M}_{\rm BH} c^2
          \equiv \epsilon_{\rm EM} \cdot \dot{m} \cdot L_{\rm Edd}/0.1,
\end{equation}
where
\begin{equation}\label{eq:eps-EM}
\epsilon_{\rm EM} = \cases{
	0.100 \times e^{-\left({\dot{M}_{\rm disk,crit}}/{\dot{M}_{\rm BH}}\right)^4},     
		\quad\,\,\,  			 	\dot{m}  >   2.3\times10^{-2}; \cr
	0.045 \times (\frac{\dot{m}}{0.01})^{0.27},
		\quad\quad\quad\quad\quad\,\,  		2.3\times10^{-2} > \dot{m}  >   9.4\times10^{-5}; \cr
	0.200 \times (\frac{\dot{m}}{0.01})^{0.59},     
		\quad\quad\quad\quad\quad\,\,  		\dot{m}  <  9.4\times10^{-5}.    
}
\end{equation}
Given the AGN luminosity, we set the Compton temperature $\Tx$ as follows 
(\textcolor{black}{cf. Equation \ref{eq:comptonization};} \citealt{sazonov_radiative_2005,xie_radiative_2017}),
\begin{equation}\label{eq:compton-temperature}
\Tx = \cases{
2.5\times10^7~~ {\rm K},     \quad\quad\quad\quad\,  l  >   0.02; \cr
1.0\times10^8~~ {\rm K},     \quad\quad\quad\quad\,  l \leq 0.02.    }
\label{compton-temperature}
\end{equation}

Finally, we ``feedback" this information to the radiative heating/cooling 
terms instantaneously, and to inner boundary conditions after 
a wind ``travel'' time $\sim R_{\rm in}/\bar{v}_w$. We assume the AGN wind is 
of a bipolar configuration, and weight the wind mass flux by $\cos^2(\theta)$. 
Provided the AGN luminosity and spectrum temperature, we are able to calculate 
the radiative heating/cooling $H-C$, and the radiation pressure due to both 
absorption and scattering are evaluated, respectively,  
\begin{equation} \label{eq:radiation-pressure}
	(\nabla p_{\rm rad})_{\rm abs} = -{H\over c},  \quad\quad 
	(\nabla p_{\rm rad})_{\rm es} = -\frac{\rho \kappa_{\rm es}}{c}
							\frac{\Lbh}{4\pi r^2}.
\end{equation}
As usual, $\kappa_{\rm es}$ is the opacity due to electron scattering.
In this way, we {can simulate} the galaxy evolution on a
cosmological timescale of 12 Gyr with AGN feedback.

\subsection{Initial and Boundary Conditions} \label{sec:numerical-setup}
To finalize the model setup, we specify the initial and boundary conditions 
in this section.
For the initial conditions,  we assume there is some remnant ISM with 1/24 
of the total stellar mass. The initial ISM is spread in the whole 
computational domain with the same density and velocity profiles as
the stellar population, and the gas is thermalized to local Virial 
temperature due to the stellar velocity dispersion and SNe Ia heating.   

At the (radial) inner and outer boundaries, we set the outflow boundary 
conditions (zero gradients, \citealt{stone_zeus-2d:_1992}). 
The AGN wind feedback is implemented by injecting mass/momentum/energy 
accordingly into the innermost 3 layers of active (radial) cells just 
next to the inner boundary. The same technique is also used to implement 
the CGM infall at the outer boundary of our computational domain.
To avoid artificial mass sources, we do not allow inflow at the outer boundary
(other than the prescribed CGM infall),
nor outflow at the inner boundary (aside from the designed BAL winds), 
and we set axial reflecting boundary condition at the azimuthal boundaries.

\begin{figure*}[htb]
\centering
\includegraphics[width=0.75\textwidth]{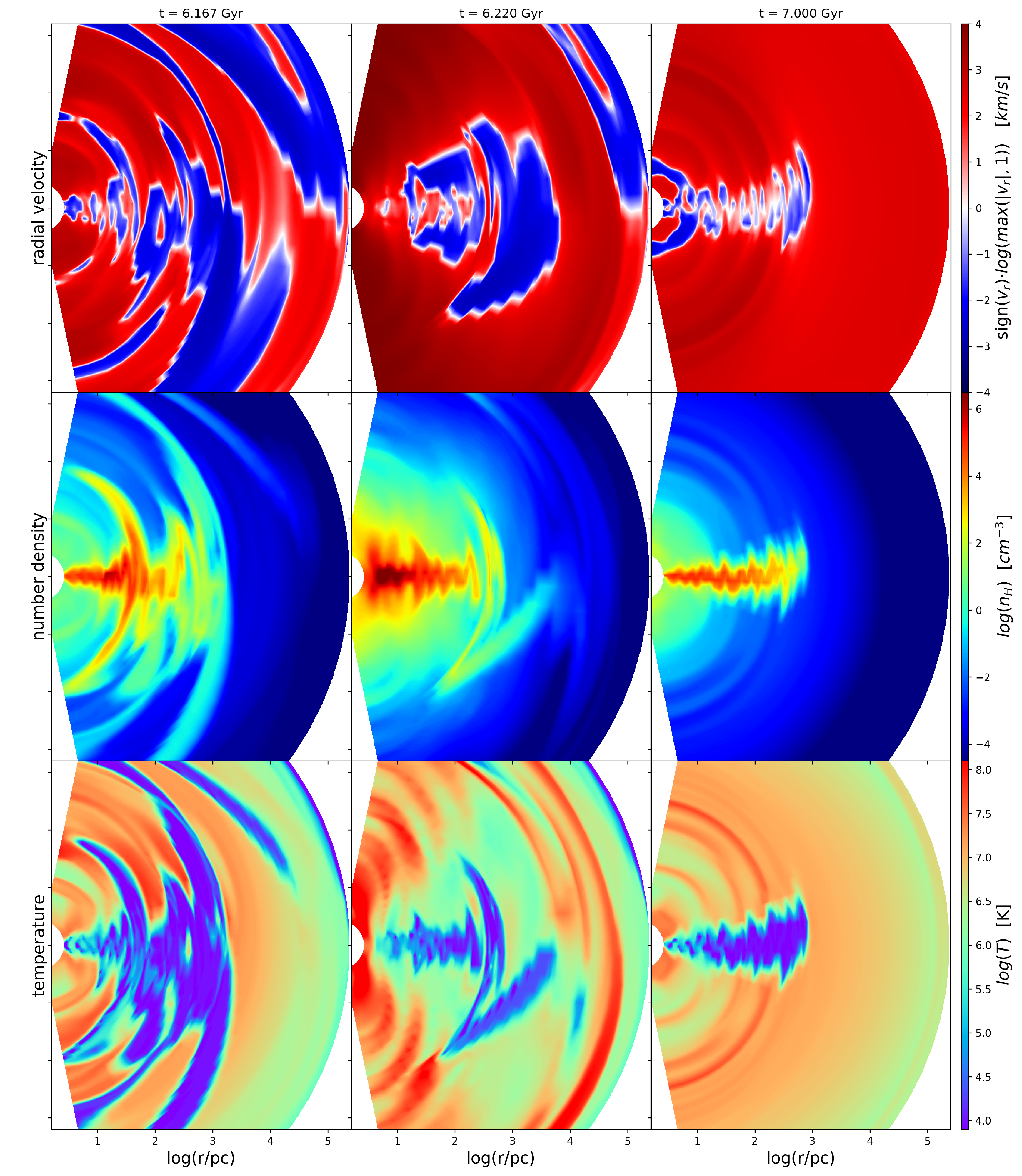}
\caption{Hydrodynamical properties of the circumnuclear disk during an AGN/star formation burst. 
From top to bottom, it shows the radial velocity (outflow in red, inflow in blue), 
number density and temperature, respectively.
From left to right, it shows the results at $t=6.167$ Gyr (just before the burst),
$t=6.22$ Gyr (during the burst), and that at $t=7.0$ Gyr (after the burst), respectively.  
Note standing shocks due to infalling gas above/below the cool central disks.}
\label{fig:cold-disk-properties}
\end{figure*}

\section{Results}
The physics ultimately driving the black hole feeding and feedback is 
the cooling flow (cf. \citealt{fabian_cooling_1994}). 
As the ISM content is continuously enriched by the 
stellar mass loss (mainly from AGB stars) and by the CGM infall 
(mass accretion from the cosmic web), it is subject to strong 
radiative cooling. As the cooling rate is proportional to density squared, 
when the ISM density increases, the cooling timescale would eventually 
become comparable to or even shorter than the local dynamical timescale, 
then a cooling flow is triggered.

As the ISM partially losses its thermal pressure, it will collapse onto 
the galaxy center and form a circumnuclear disk because of the angular 
momentum barrier. The dynamical timescale of the disk is much longer than 
the free-fall timescale, so it allows mass to accumulate in the disk and to 
be cooled down catastrophically.

The cold disk would be extremely over-dense, and it is subject to gravitational 
instability and to star formation. Meanwhile, the spiral waves, 
as a consequence of the gravitational instability, would also help to 
transfer angular momentum so as to allow mass to be accreted onto the galaxy center. 
It turns out most of the cold gas would be consumed by star formation 
on its way to the supermassive black hole.

Finally, some of the gas would be accreted by the supermassive black hole, 
which lights up as an AGN (or even quasar). Consequently, strong AGN feedback 
is capable of altering all the processes above, to drive galactic outflow, 
and to regulate the black hole accretion itself by injecting huge amounts 
of energy and momentum (in terms of both radiation and wind) back to 
its host galaxy. The cooling flow will be quenched after large outbursts 
until another cycle starts over again.

\begin{figure}[htb]
\centering
\includegraphics[width=0.525\textwidth]{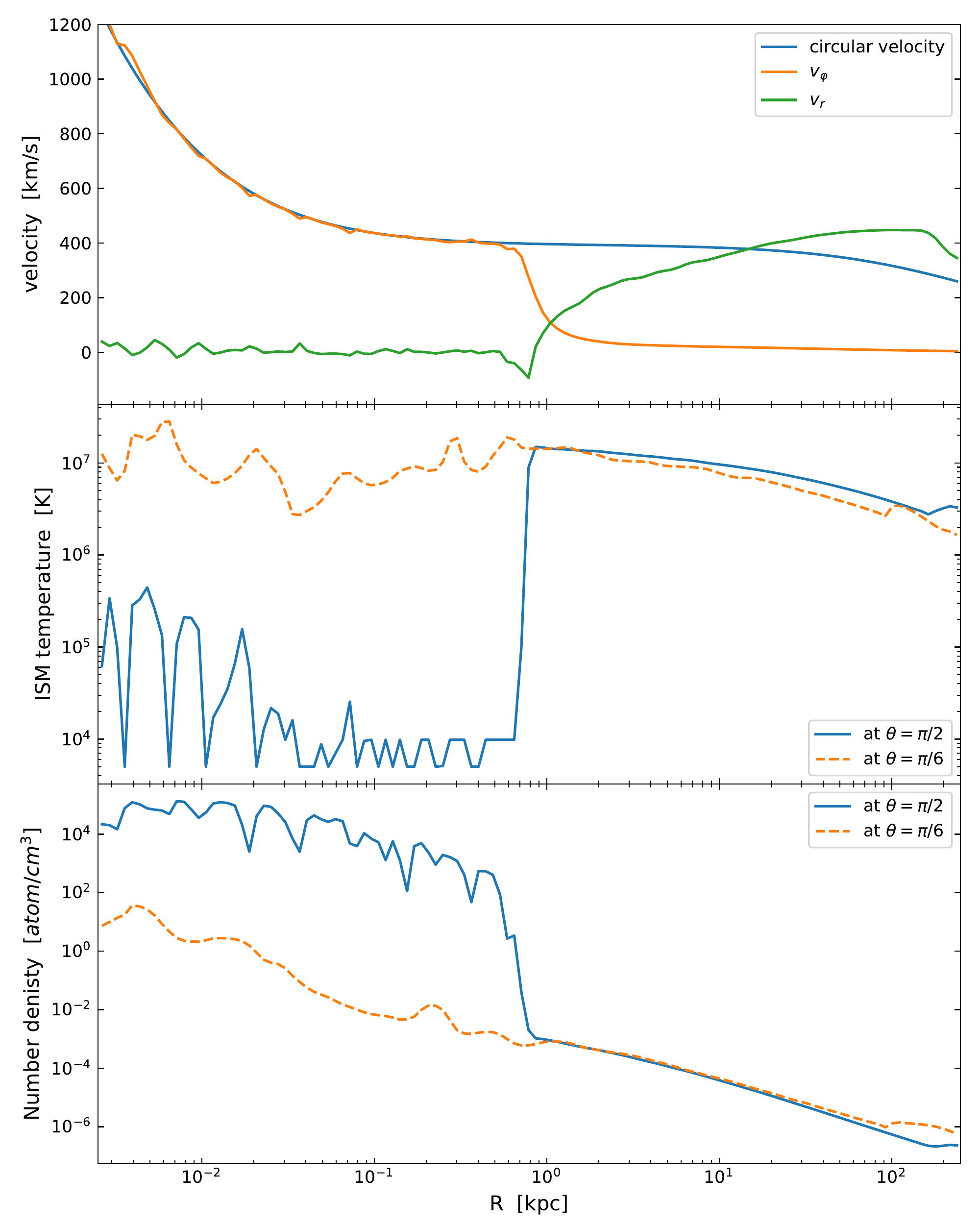}
\caption{Circularization of the circumnuclear disk. 
The ISM velocity, temperature and density profiles on ($\theta=\pi/2$, solid lines) 
and off ($\theta=\pi/6$, dashed line) the disk (at $t=12$ Gyr) are shown in the upper, middle and lower panels, 
respectively. The ISM in the circumnuclear disk simply hits the temperature floor, 
and is solely supported by rotation against gravity.}
\label{fig:cold-disk-details}

\vspace{0.01\textwidth}

\includegraphics[width=0.4\textwidth]{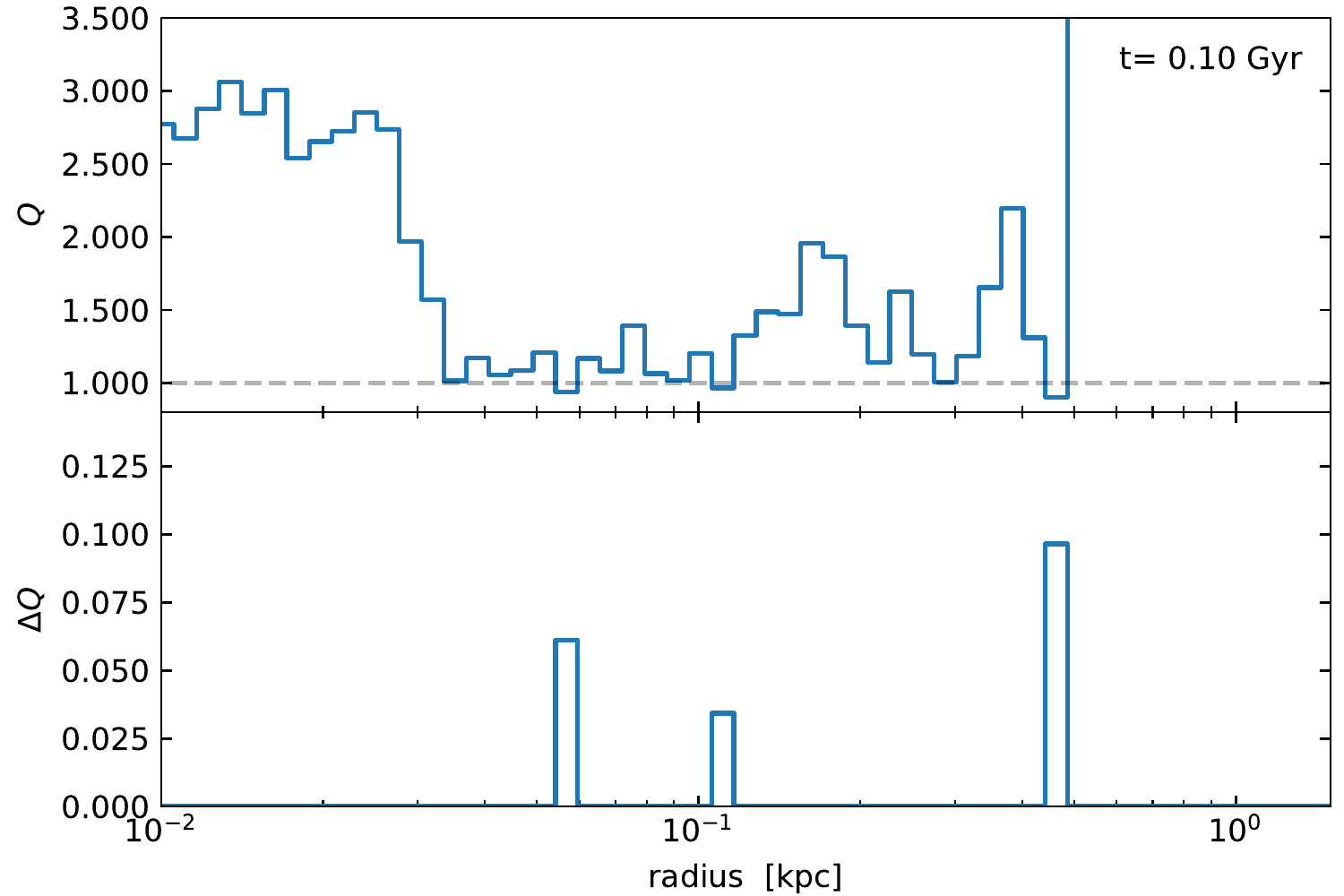}
\includegraphics[width=0.4\textwidth]{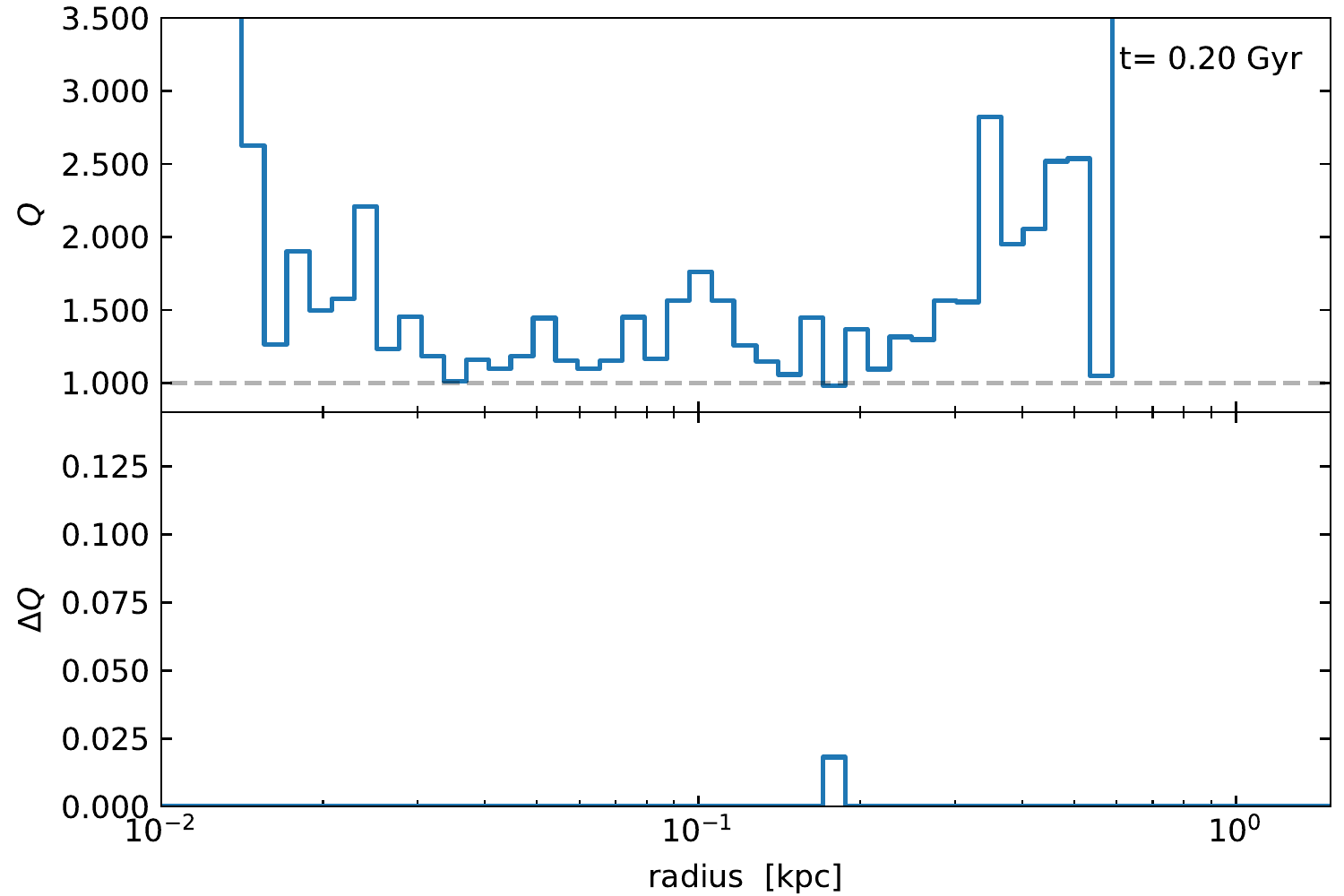}
\includegraphics[width=0.4\textwidth]{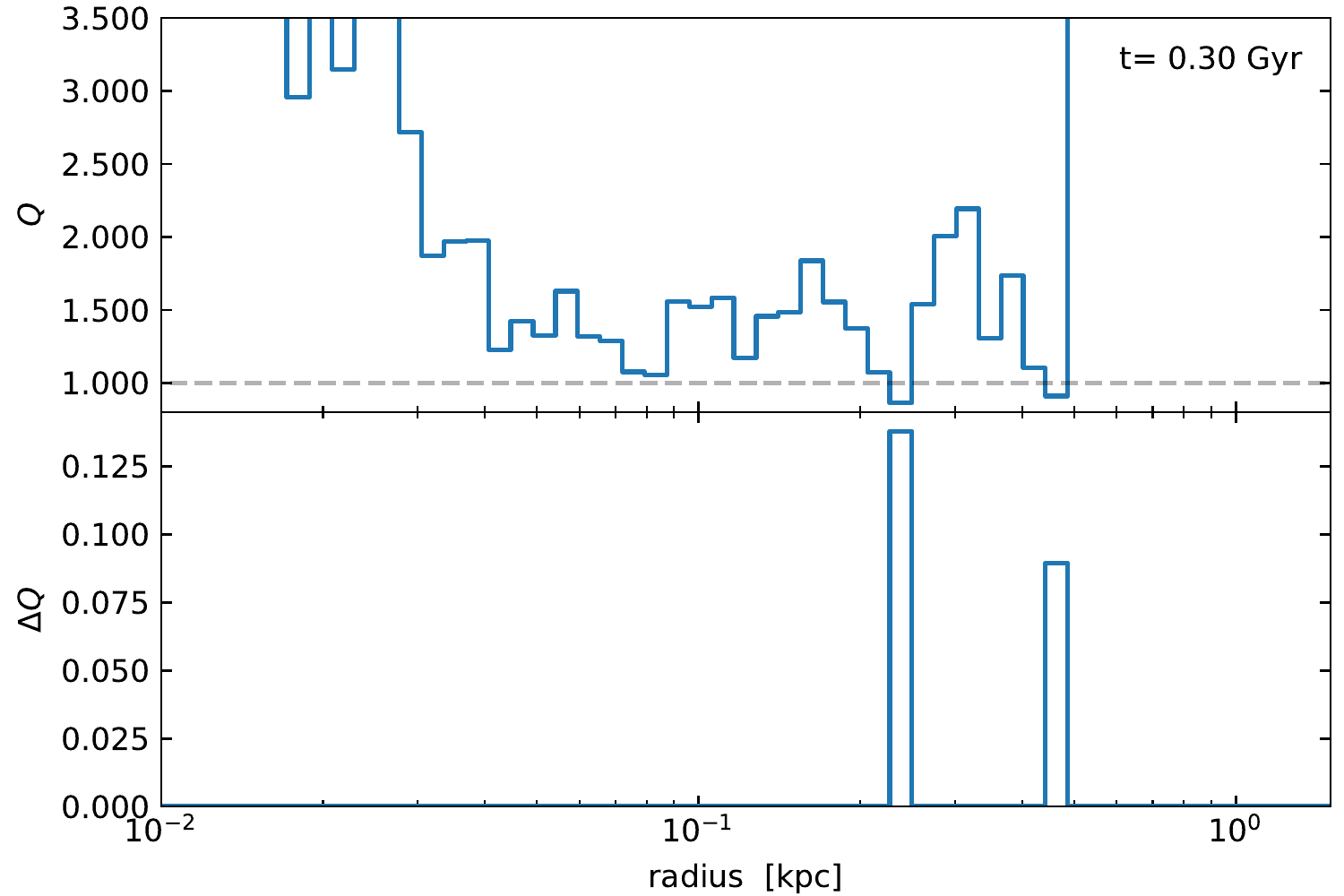}
\includegraphics[width=0.4\textwidth]{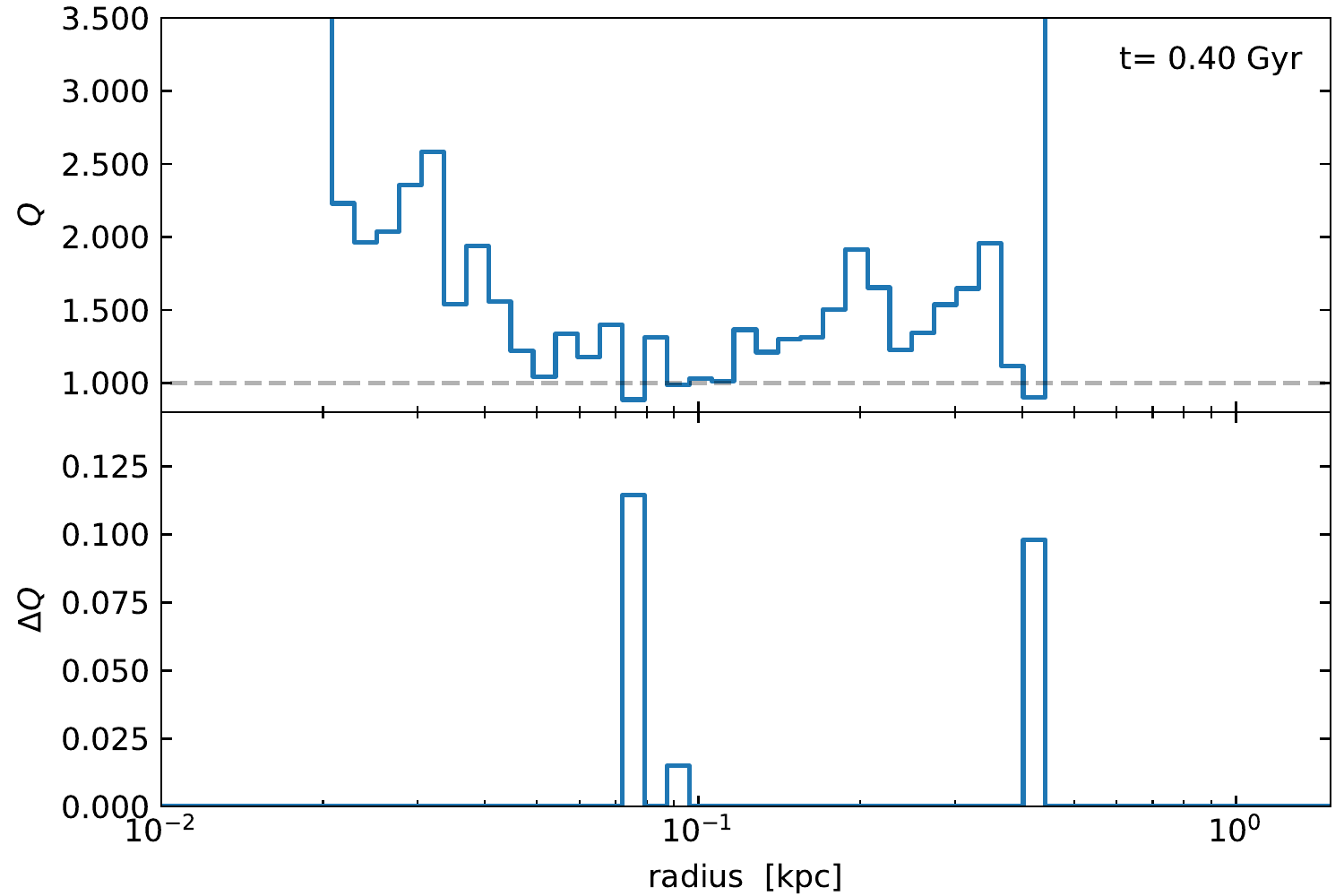}
\caption{\textcolor{black}{The Toomre $Q$ parameter. In the four panels, we plot the results on the equatorial plane at a selection of representative times, i.e., $t=0.1, \ 0.2,\ 0.3,\ 0.4$ Gyr, respectively. In each panel, it shows the Toomre $Q$ parameter in the top sub-plot and $\Delta Q\equiv \max(1-Q,0)$ in the bottom sub-plot. We can see that the Toomre instability occurs in individual disk rings.}}
\label{fig:Toomre-Q}
\end{figure}

\subsection{Toomre Instability in The Cold Circumnuclear Disk}
In Figure \ref{fig:cold-disk-properties} we plot the ISM profiles of 
radial velocity, density and temperature during an AGN/star formation burst. 
Note the bi-conical outflow in the middle vertical panels. 
We can clearly see some ripples induced by AGN feedback 
and a cold disk of size $\sim1$ kpc 
wiggling and sitting in the equatorial plane. More details of the cold disk 
are shown in Figure \ref{fig:cold-disk-details}. We can see that the disk 
is cooled down to $\sim10^4$ K (which simply hits the temperature floor 
of our numerical model). Such a temperature is far below the local 
Virial temperature $\sim10^7$ K (see the orange line in the middle panel of Figure \ref{fig:cold-disk-details}). 
It turns out the cold disk is fully supported 
by rotation against the gravity, as shown in the upper panel, the rotation 
profile of the disk fits perfectly with the analytical circular velocity 
derived from the total gravitational potential 
(see Equation \ref{eq:circular_velocity}). 
The jump in pressure at the upper and lower surfaces of the disk is balanced by 
ram pressure of the infalling gas.

\begin{figure*}[htb]
\centering
\includegraphics[width=0.6\textwidth]{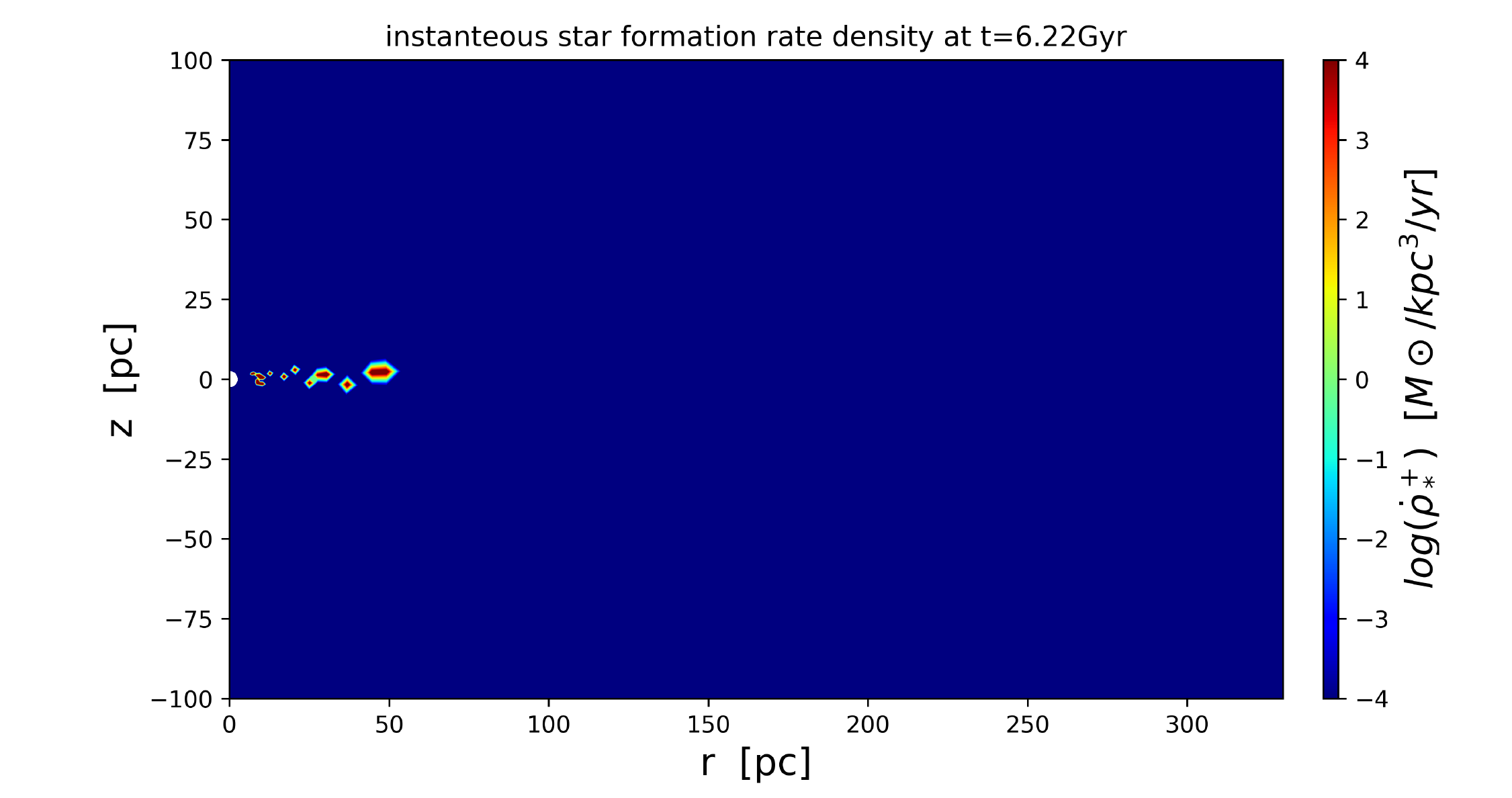}
\includegraphics[width=0.6\textwidth]{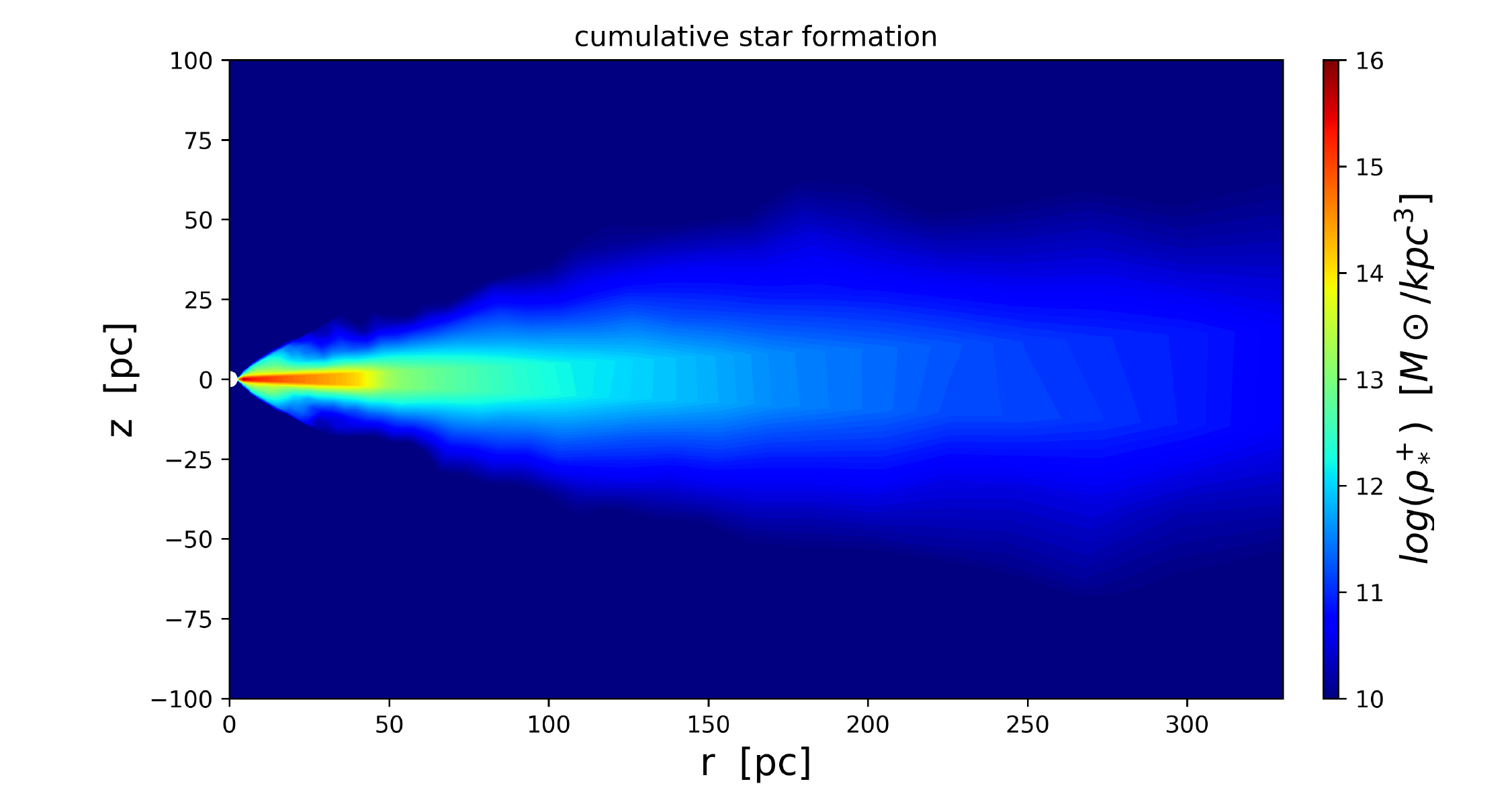}
\caption{Spatial distribution of star formation. 
Upper panel: the instantaneous star formation rate density in our fiducial 
	at $t=6.22$ Gyr (instantaneous star formation occurs in individual disk rings).
Lower panel: cumulative star formation at the end of the simulation 
	(Most of star formation occurs in the cold disk). }
\label{fig:star-formation-spatial-distribution}

\vspace{0.01\textwidth}

\includegraphics[width=0.55\textwidth]{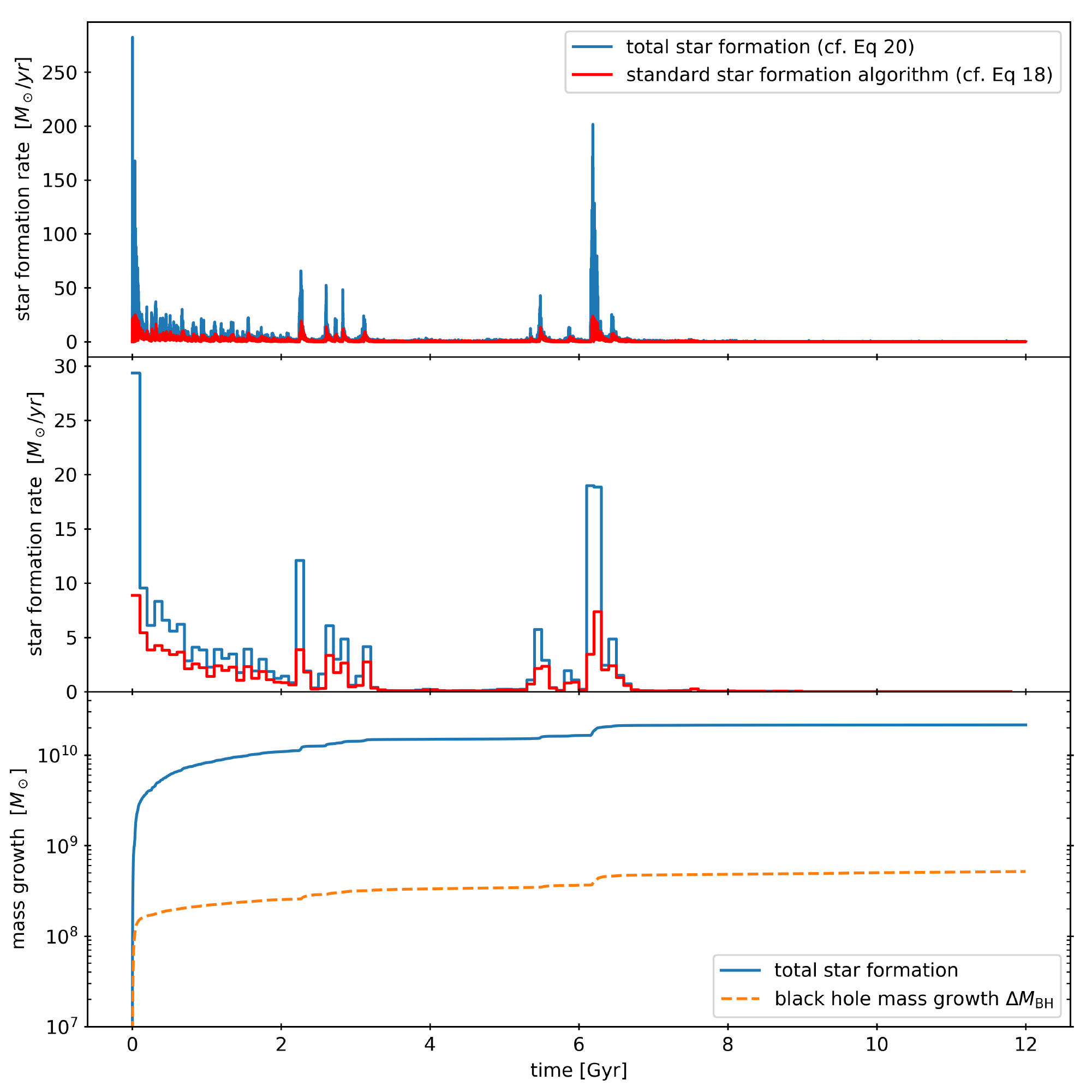}
\caption{Total star formation integrated over the whole galaxy. 
Top panel: star formation rate versus time;
Middle panel: same as the top panel except that the data points are
   binned and averaged over equal time intervals of $\Delta t =10^8$ year;
Bottom panel: cumulative star formation before given time (blue line). 
The black hole mass growth $\Delta M_{\rm BH}$ (orange dashed line), 
which is synchronous with the star bursts, is also plotted for reference.}
\label{fig:star-formation-history}
\end{figure*}

As the ISM cools down and continuously falls onto the disk
(cf. Figure \ref{fig:cold-disk}, left panel), its surface density increases. 
The disk can be finally over-dense and becomes Toomre unstable 
in some {individual rings}, 
\textcolor{black}{though the fraction of the unstable disk rings
is extremely low (as shown in Figure \ref{fig:Toomre-Q}
in which we plot the profiles of the Toomre Q parameter 
at a selection of representative times).}
As described in \S\ref{Q-unstable} and \ref{sec:star-formation}, 
the Toomre instability is capable of transferring mass inward, 
and will trigger star formation in the meantime 
(see Figure \ref{fig:star-formation-spatial-distribution}, upper panel).
As a result, the surface density decreases and the disk rings will be stabilized again.
Because of the surface density threshold of the Toomre instability (Equation \ref{eq:Toomre-Q}),
such processes are always bursty.

\begin{figure}[htb]
\centering
\includegraphics[width=0.625\textwidth]{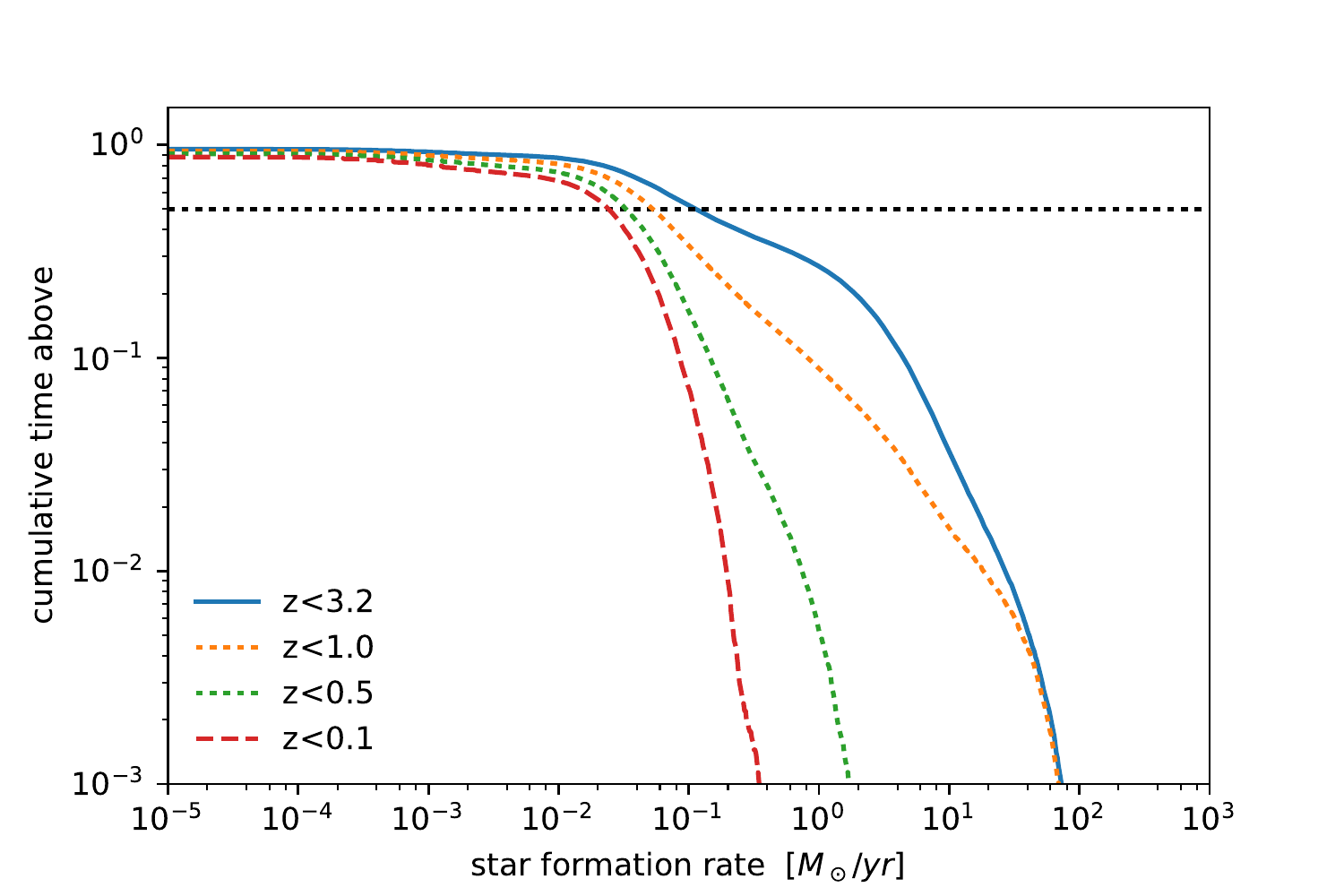}
\caption{Duty cycle of star formation, i.e. percentage of cumulative time above given star formation rate. The horizon dotted line represents a fixed duty cycle of $50\%$. 
}\label{fig:star-formation-duty-cycle}

\vspace{0.03\textwidth}

\includegraphics[width=0.54\textwidth]{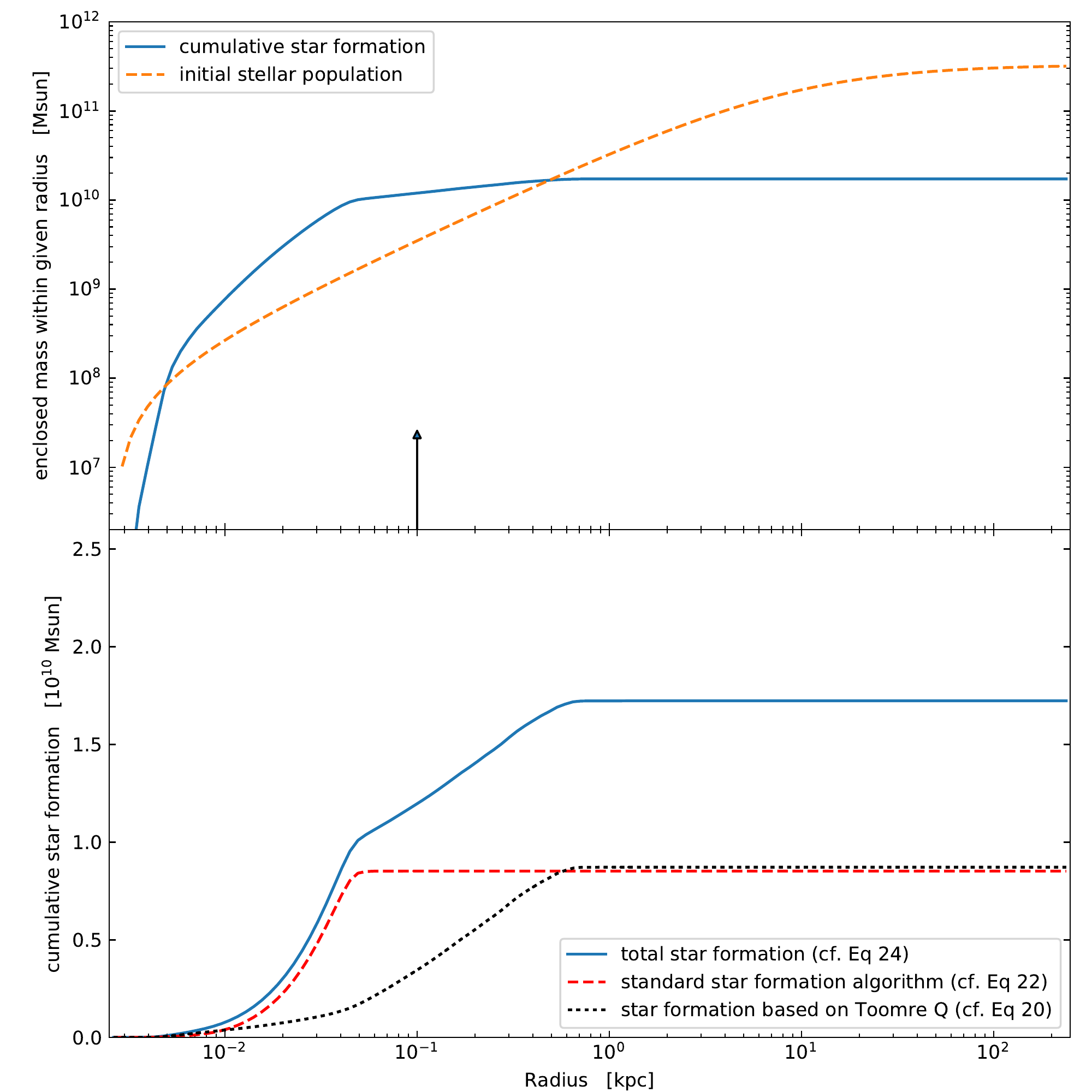}
\caption{Total mass of the new (blue line) and old (orange dashed line) stellar populations enclosed by given radii. Most of the star formation occurs within $r \le 3$ kpc. The mass of the newly formed stars is larger than the old stellar population around $r\leq1$ kpc. The vertical arrow shows approximately the radius of influence of the black hole ($\sim100$ pc). Inner boundary of the simulations is 2.5 pc.
} \label{fig:stellar-mass-profile}
\end{figure}

\begin{figure}[htb]
\centering
\includegraphics[width=0.55\textwidth]{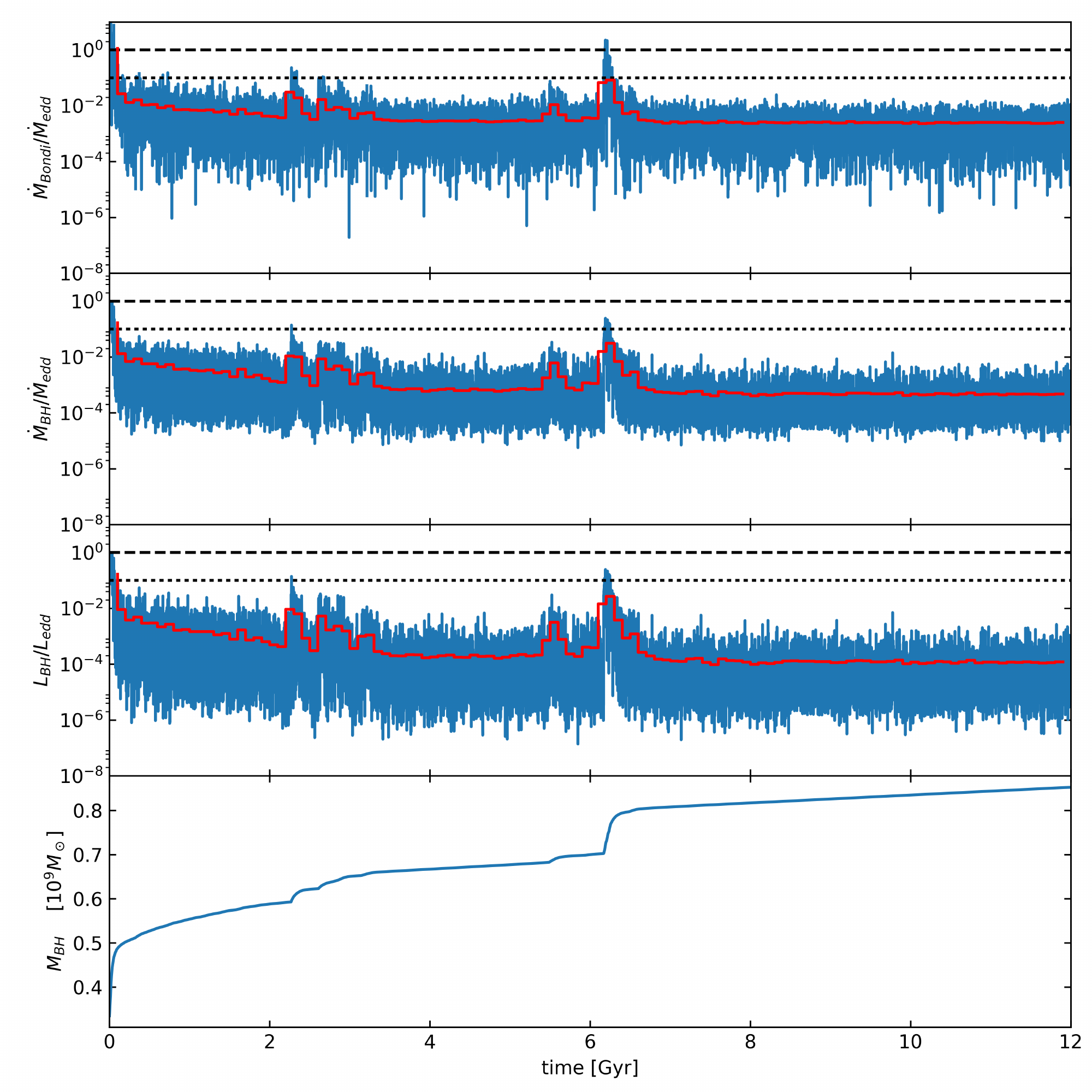}
\caption{AGN activities and black hole mass growth.
From top to bottom, the blue lines present, respectively,  
(1) the mass inflow at inner boundary,
(2) the mass accretion rate onto the event horizon of the central supermassive black hole, 
(3) the bolometric luminosity of the AGN, and
(4) the black hole mass.
The red lines are the same as the blue lines except that the data points are
   binned and averaged over equal time intervals of $\Delta t =10^8$ year.
The mass flow rate and the AGN luminosity are normalized by the Eddington values (determined by the instantaneous black hole mass). The black dashed and dotted lines represent constant Eddington ratios of 1.0 and 0.1, respectively. 
}\label{fig:black-hole-accretion}
\vspace{0.01\textwidth}
\includegraphics[width=0.55\textwidth]{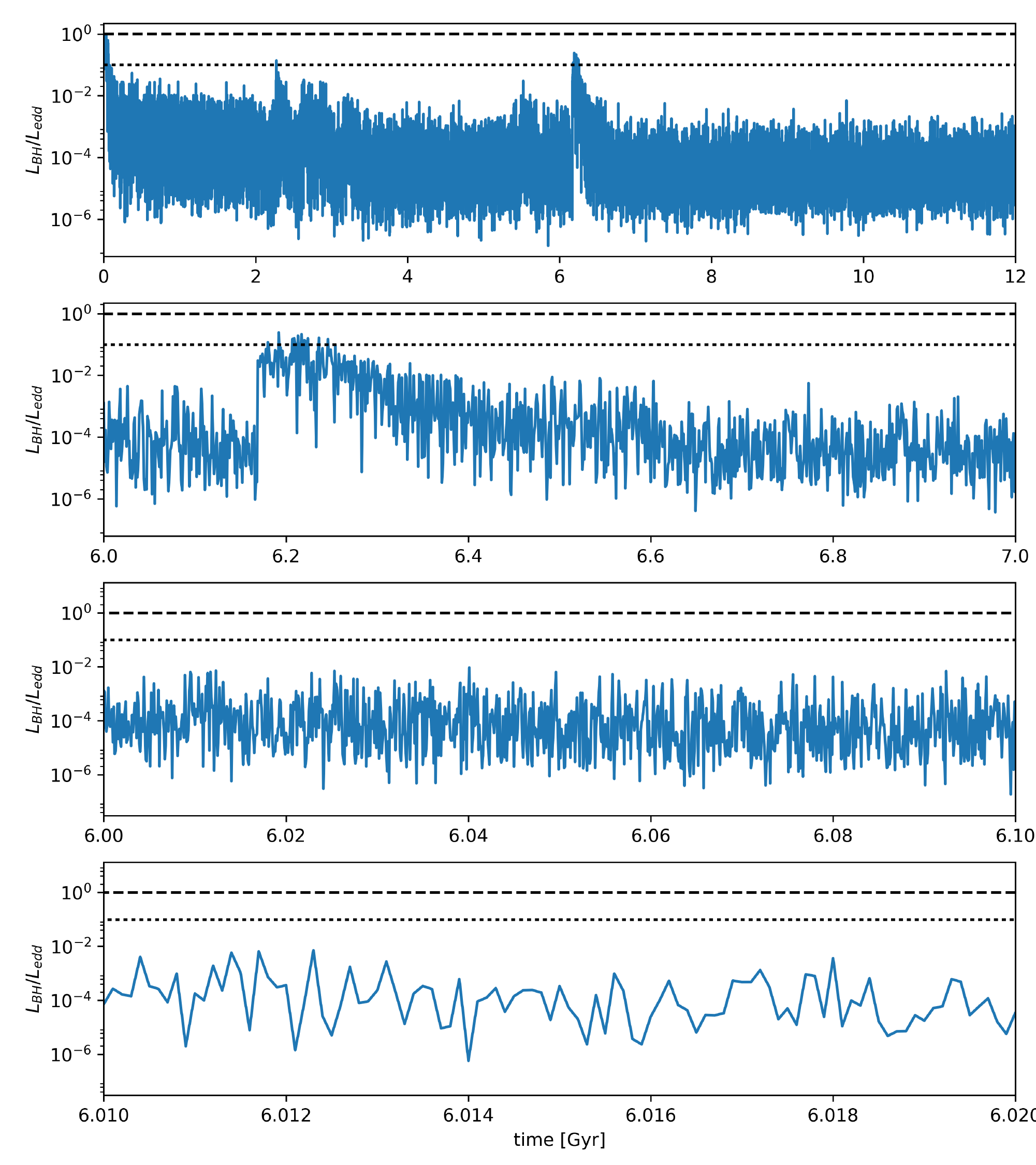}
\caption{AGN bolometric luminosity as in Figure \ref{fig:black-hole-accretion} (zoomed).}
\label{fig:agn-light-curve-zoomed}
\end{figure}

\subsection{Star Formation History}
All star formation occurs in the cold disk (as shown in the lower panel of 
Figure \ref{fig:star-formation-spatial-distribution}). More precisely, most of 
the star formation takes place in individual disk rings where it is subject to 
Toomre instability (as shown in the upper panel of 
Figure \ref{fig:star-formation-spatial-distribution}). So, it is intrinsic that 
the new stars will be born in bursts. The star formation history is shown 
in Figure \ref{fig:star-formation-history}.  In Figure \ref{fig:star-formation-duty-cycle}
we analyze the duty cycle of star formation,  i.e. percentage of cumulative time 
above given star formation rate.
In Figure \ref{fig:stellar-mass-profile}, we show the enclosed mass 
of the cumulative star formation and compare it to the initial stellar mass 
profile.

We can see that most of the star formation occurs in the circumnuclear 
disk of a size $\le 1$ kpc during the bursts in the early stage evolution, 
which is in agreement with 
recent observations. For example, \citet{tadaki_gravitationally_2018}
observed the starburst galaxy AzTEC-1 (z=4.3) using ALMA. They
found that a large fraction of stars is formed in the central 1 kpc region, 
plausibly in a gravitationally unstable gas disk.  Such an observational
phenomena matches very well with the early bursts 
that we find in our simulations.

We note that the star formation rate in our simulation is actually low in most of time, 
especially in the late stage ($\le 0.05 M_\odot/{\rm yr}$), 
and it tends to be located on the very central ($r<25$ pc) regions 
(\citealt{tan_star-forming_2005}). 
The total star formation is $\sim$ few percents of the initial stellar mass. 	
In Figure \ref{fig:stellar-mass-profile}, 
we can also see that the new star mass could become larger than 
the initial stellar mass at $r\le1$ kpc, which could mildly alter 
the gravity profile in the central region. 
\textcolor{black}
{However, for simplicity, we don't consider the gravity of the new stars in this paper. 
We leave it to our future work in which we will consider a time-dependent 
galaxy dynamics model.}

\subsection{AGN Activities and Black Hole Mass Growth} \label{sec:agn-feedback}
\begin{figure}[htb]
\centering
\includegraphics[width=0.55\textwidth]{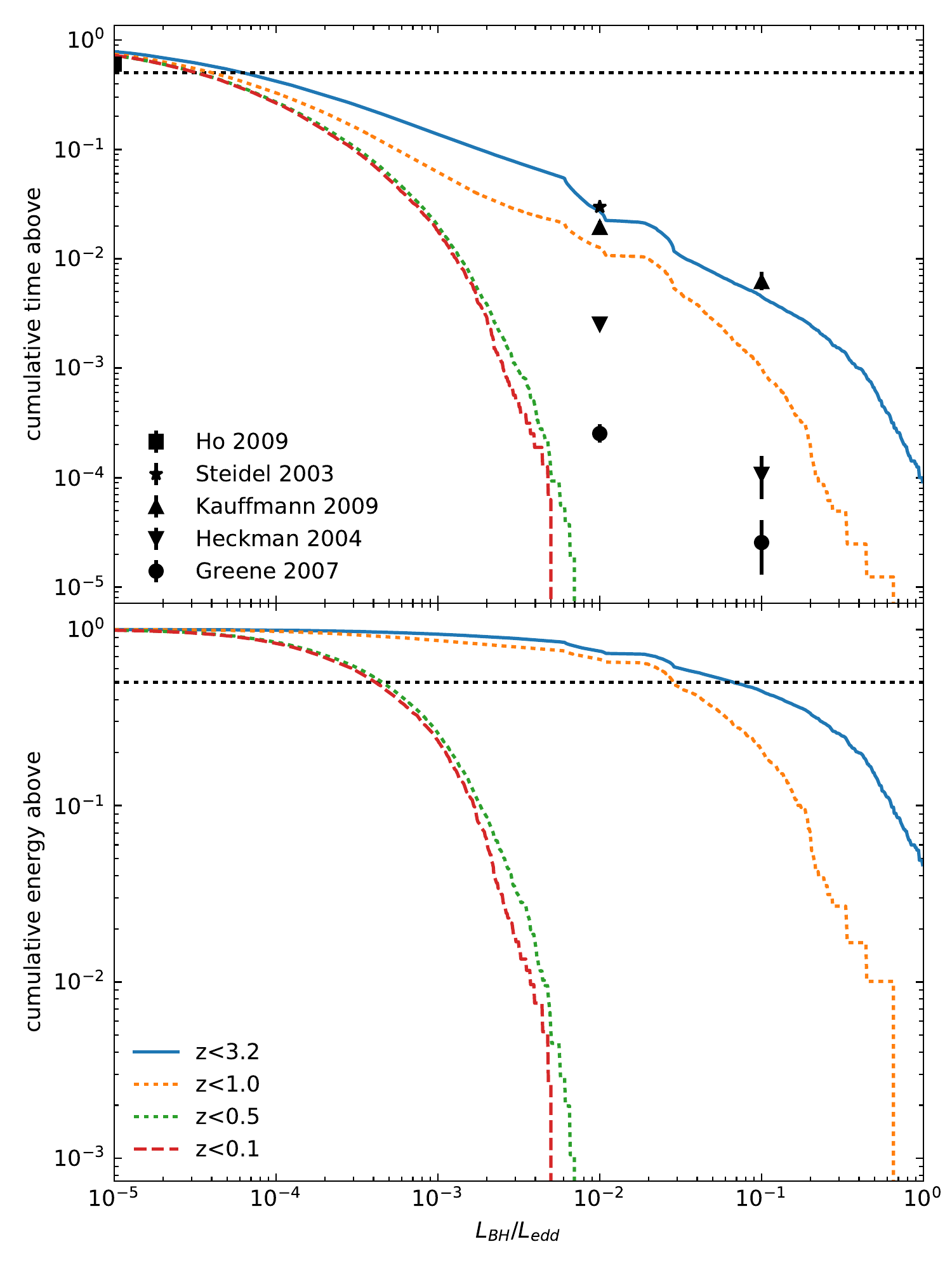}
\caption{AGN duty cycle, i.e. percentage of cumulative time (upper panel) and cumulative energy (lower panel), above given Eddington ratio. The horizon dashed lines represent a fixed duty cycle of $50\%$. 
The lines in colors are the simulation results below given red shift.
The points are observational constraints. The squares, circles, and upward- and downward- pointing triangles are from Ho (2009), Greene \& Ho (2007), Kauffmann \& Heckman (2009), and Heckman et al. (2004), respectively, which are all compiled from low-redshift observations.
The star is a constraint compiled from high-redshift observations by Steidel et al. (2003).
}\label{fig:agn-duty-cycle}
\end{figure}

Similar to star formation, the black hole accretion is also bursty, 
as it is driven by the same physical processes, i.e., the Toomre instability. 
In Figure \ref{fig:black-hole-accretion}, we plot the black hole accretion 
history, from the top to bottom panel, it shows the mass inflow rate 
via the inner boundary, 
the mass accretion rate down to the black hole 
event horizon, the consequent AGN bolometric luminosity, and the black hole 
mass growth, respectively (more details of the AGN light curve can be found in 
Figure \ref{fig:agn-light-curve-zoomed}). 
In Figure \ref{fig:agn-duty-cycle} we analyze 
the AGN duty cycle in terms  of cumulative energy/time when the AGN luminosity 
(the Eddington rate) is above given values. We can see that it agrees well 
with the Soltan argument, i.e.,  the AGN spends most of its life time 
at very low luminosity, while emitting most of its energy when 
it is at high luminosity \citep{soltan_masses_1982}.


\subsection{Overall Mass Budget} \label{sec:mass-budget}
\begin{figure}[htb]
\centering
\includegraphics[width=0.7\textwidth]{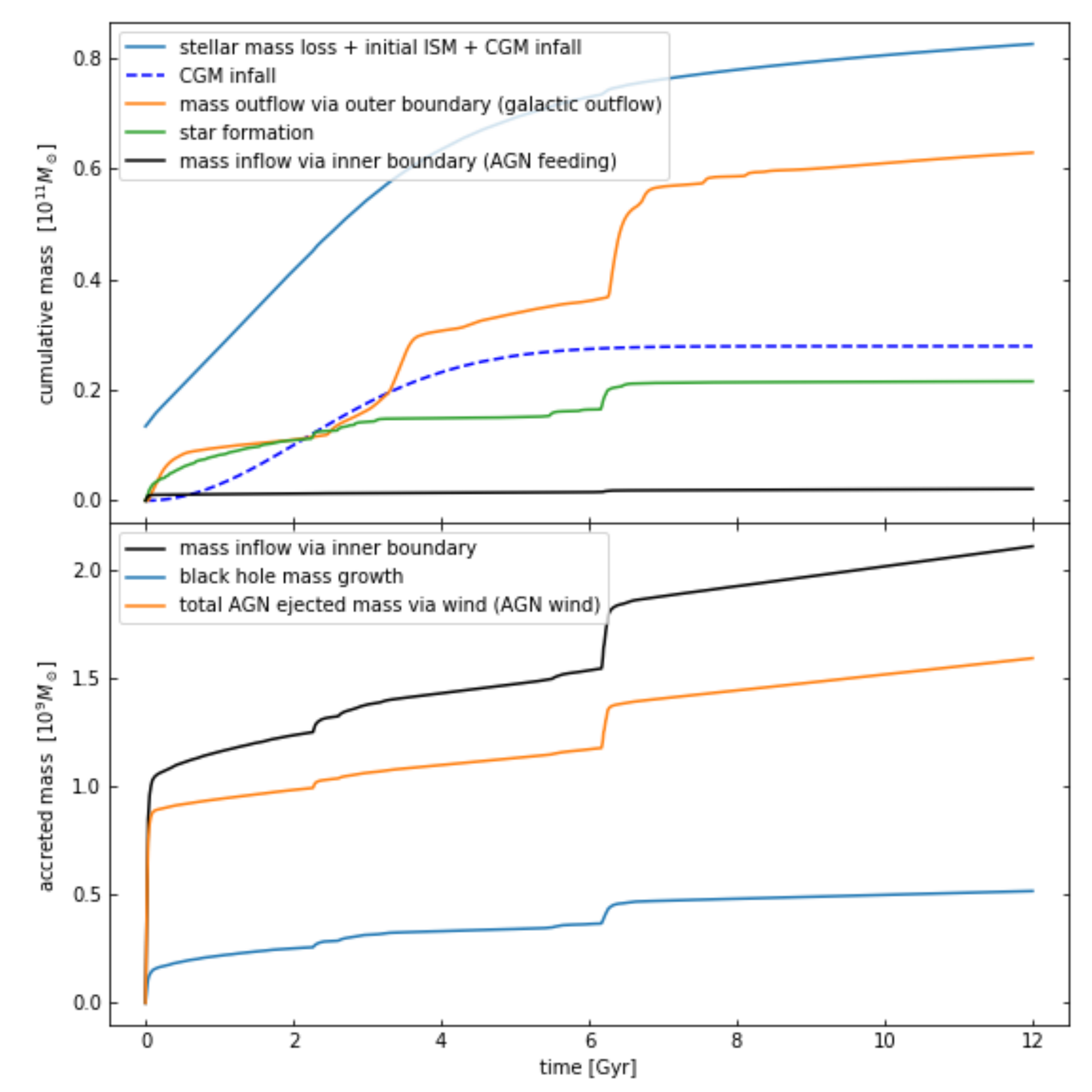}
\caption{Mass budget, of the total ISM content (upper panel) and of the black hole feeding (lower panel),  over the cosmological evolution. The upper panel shows: (1) the cumulative mass of the ISM sources (blue solid line; including the initial ISM remnant, the CGM infall (see the blue dashed line) and the stellar mass loss), (2) the cumulative mass of the galactic outflow that escaped from the outer boundary (orange line; which dominates the mass budget), (3) the cumulative mass of star formation (green line), (4) the cumulative mass inflowing via inner boundary (i.e. black hole feeding, black line). 
The lower panel shows: (1) the cumulative mass fed to the supermassive black hole (black line), (2) the black hole mass growth $\Delta M_{\rm BH}$ (blue line), and (3) the total mass injected by the AGN (via AGN wind feedback, orange line).}
\label{fig:mass-budget}

\vspace{0.01\textwidth}

\includegraphics[width=0.65\textwidth]{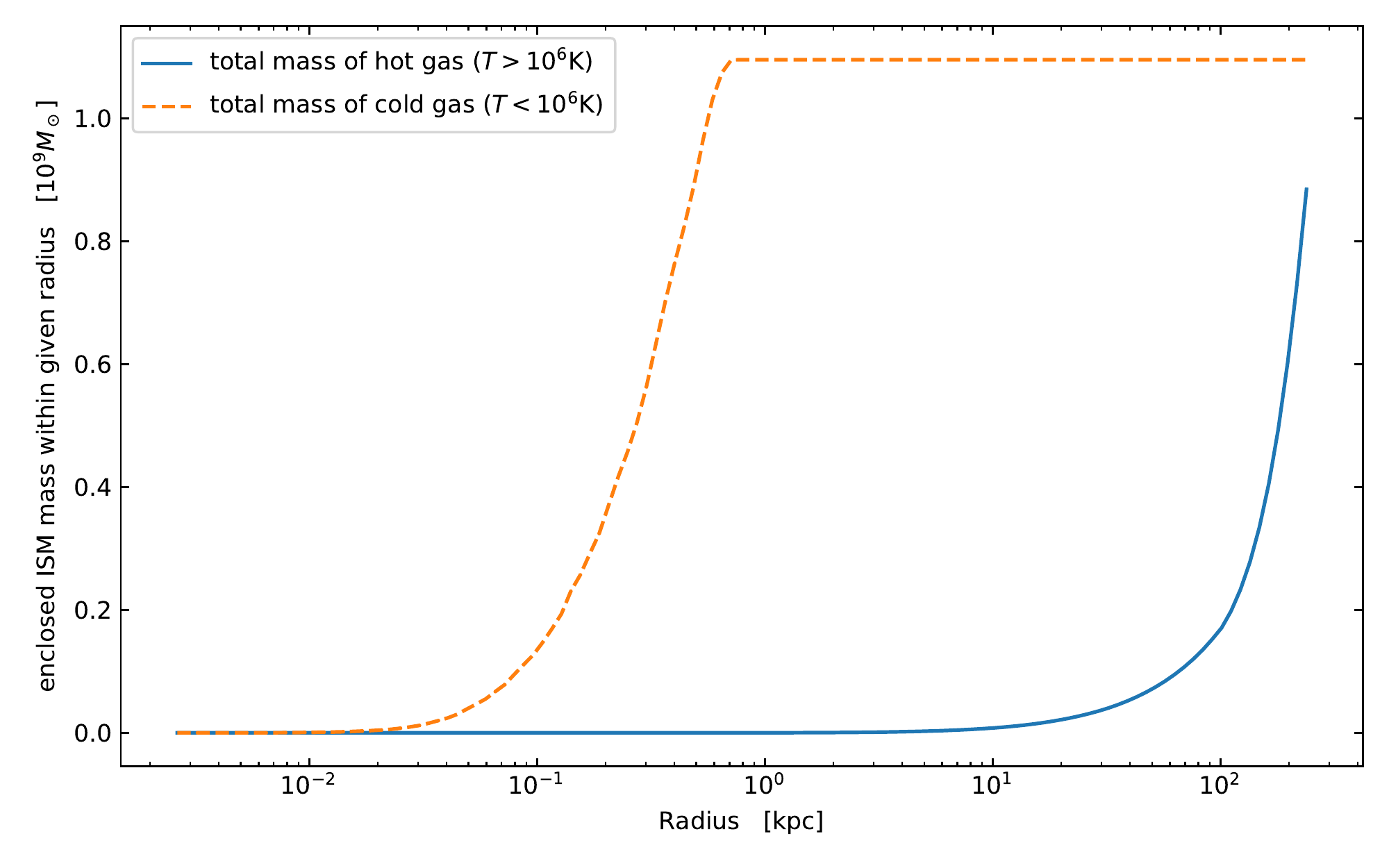}
\caption{The total ISM mass remaining, in forms of hot gas ($T>10^6$ K; blue line) and cold gas ($T<10^6$ K; orange dashed line), at the end of the fiducial run. The cold gas is mainly in the circumnuclear disk within $r\le1$ kpc, while the hot gas is mainly in the galaxy outskirts (which minimally contributes to the total ISM X-ray luminosity because of its low density).}
\label{fig:ism_mass_spacial_distribution}
\end{figure}

We track the overall mass \textcolor{black}{budget} in Figure \ref{fig:mass-budget}. 
As the mass source comes from (1) stellar mass loss, (2) CGM infall, 
and (3) the initial ISM content, we can see that most of the gas is expelled 
out of the galaxy, especially during the AGN bursts. 
One quarter of the gas is consumed by star formation. 
Only a small fraction is accreted by the supermassive black hole. 
The rest remains in the galaxy, of which some is placed within the cold disk 
and the other is in the form of hot phase ISM (which is capable of emitting X-rays). 
In the lower panel, we can see that most of the mass inflow onto the galactic center 
(black line) is ejected as winds (orange line) and only a small fraction is finally accreted 
by the central black hole (blue line).

\begin{figure}[htb]
\centering
\includegraphics[width=0.95\textwidth]{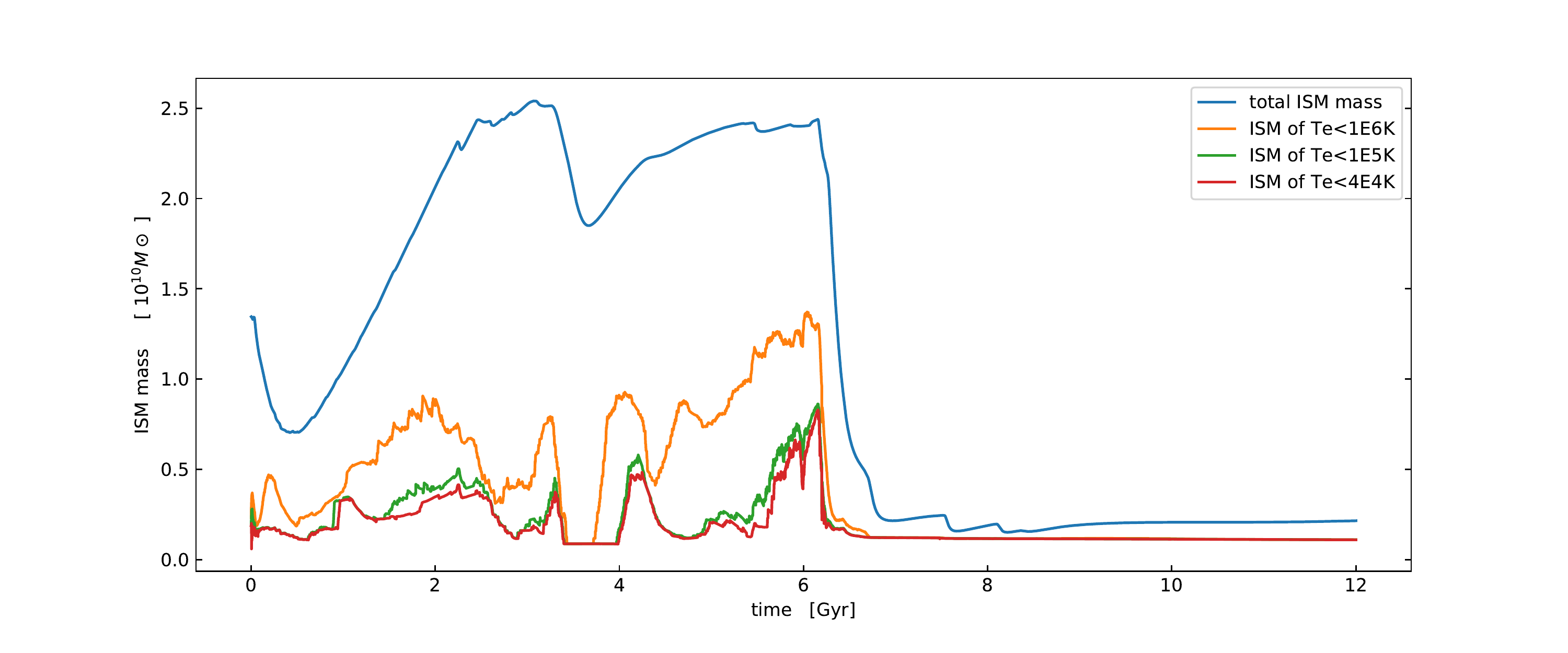}
\caption{The time evolution of the cold phase ISM. The rapid drops after the peaks are due to the Toomre instability, which drives both star bursts (cf. Figure \ref{fig:star-formation-history}) and strong AGN feedback (cf. Figure \ref{fig:black-hole-accretion}). Note that the cold gas with $T<10^6$ K is mainly in the disk. }\label{fig:ism-mass}

\vspace{0.01\textwidth}

\includegraphics[width=0.75\textwidth]{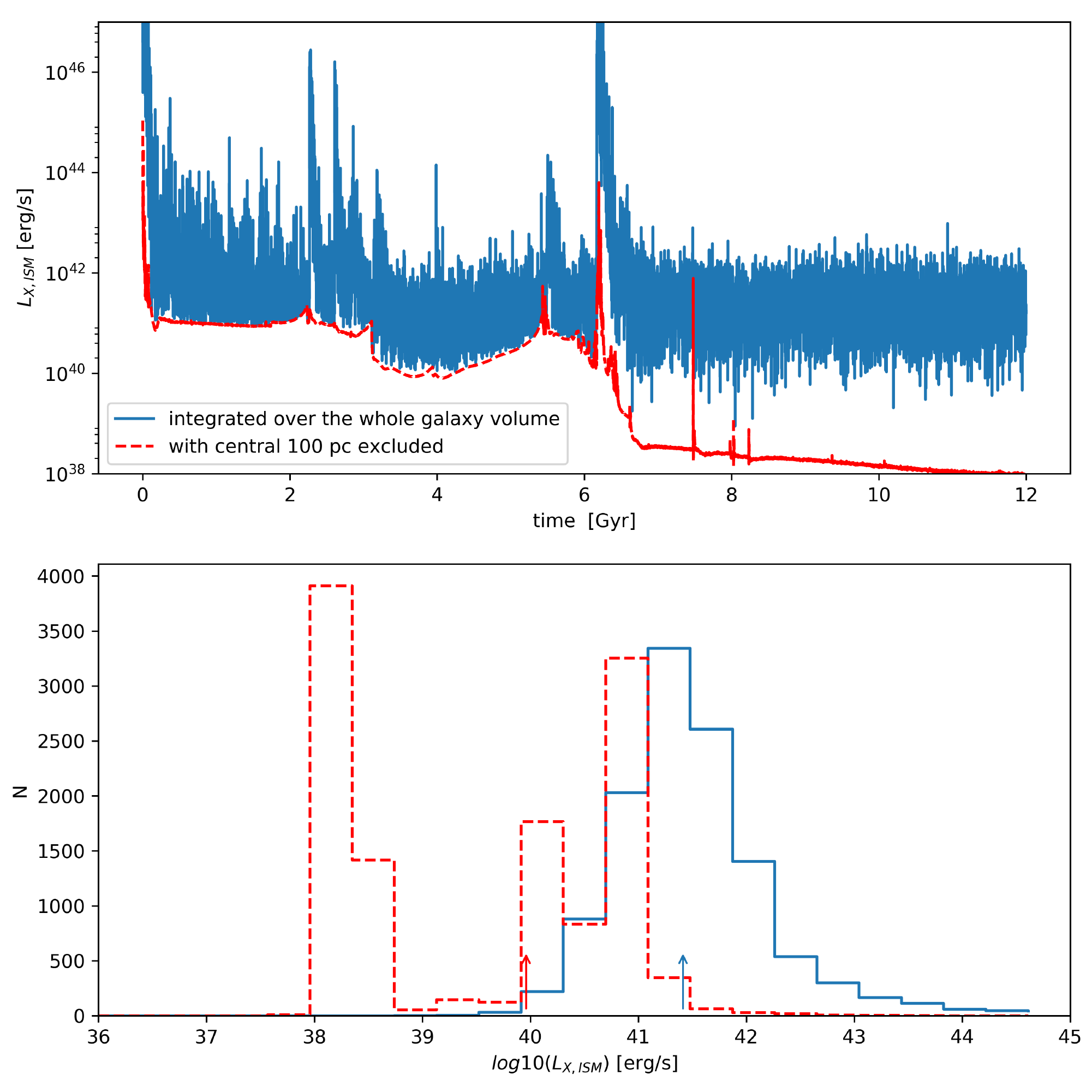}
\caption{ISM X-ray luminosity in the band of 3-8 kev. The luminosity is calculated by integrating the frequency-dependent emissivity over the whole galaxy volume (blue lines; the blue vertical arrow in the lower panel indicates its median value $L_{\rm X, ISM}=2.5\times10^{41}$ erg/s). The red dashed lines show the results that with central 100 pc excluded (the red vertical arrow in the lower panel indicates its median value $L_{\rm X, ISM}=9.1\times10^{39}$ erg/s). The atomic data needed in the calculation are extracted from the \texttt{ATOMDB} code (version 3.0.9). In the lower panel, we sample the ISM X-ray luminosity with equal time interval of 0.01 Gyr, and bin the data in logarithmic scale.}
\label{fig:ism-xray}
\end{figure}

The spatial distribution of the remaining gas is shown in 
Figure \ref{fig:ism_mass_spacial_distribution}. We can see that 
most of the hot ISM is located at the outskirts of the galaxy, 
where the density is too low to contribute to the X-ray luminosity. 
In Figure \ref{fig:ism-mass} we plot the time evolution of the ISM content, 
while in Figure \ref{fig:ism-xray} we plot the ISM X-ray luminosity. 
We see that the ISM X-ray luminosity lies in a reasonable range 
and agrees well with observations.

\section{Effects of New Physics}
 In this section, we present the results of four control models, 
 which are based on the fiducial run we showed previously, 
 but with rotation, the Toomre instability,  the $\alpha$ viscosity, 
 or the hot mode feedback turned off, respectively. 
 In this way, we try to analyze the effects of the new physics 
 we include in this paper. 
 The statistical properties of the control models are summarized in
 Table \ref{tab:control-models}.

\begin{table}[ht]
\caption{Statistical properties of the modeling galaxies}\label{tab:control-models}
\begin{center}
\begin{tabular}{cccccccccc}
\hline\hline
\multirow{2}{*}{Model \#} &\multicolumn{3}{c}{AGN duty cycle} 
& {$\Delta M_{\rm BH}$ $^d$} & { $\Delta M_{\rm w, AGN}$ $^e$} & {$\Delta M_{\rm w, Gal}$ $^f$} 
& {$\Delta M_{\star}^+$$^g$} & {$<SFR>$$^h$} & { disk size$^i$} \\
\cmidrule{2-4} & {$f_{E, l>0.1}$ $^a$} &{$l_{\rm median, E}$ $^b$} & {$l_{\rm median, t}$$^c$} 
& $M_\odot$ & $M_\odot$ & $M_\odot$ & $M_\odot$ & $M_\odot/yr$ & kpc \\
\hline
fiducial & 44.6 \% & $6.7\times10^{-2}$ & $6.3\times10^{-5}$ & $5.17\times10^8$ & $1.59\times10^9$ & $6.28\times10^{10}$ & $2.15\times10^{10}$ & $3.5\times10^{-2}$ & 0.64 \\
C1$\ \ $ & 51.1 \% & $1.1\times10^{-1}$ & $8.0\times10^{-5}$ & $1.01\times10^9$ & $3.69\times10^9$ & $7.41\times10^{10}$ & $2.06\times10^{5\ \ }$ & 0.00 & 0.00 \\
C2$\ \ $ &6.06 \% & $2.9\times10^{-3}$ & $9.2\times10^{-5}$ & $1.59\times10^8$ & $3.36\times10^8$ & $7.87\times10^{9\ \ }$& $6.86\times10^{9\ \ }$ & $4.5\times10^{-1}$ & 1.52\\
C3$\ \ $ & 77.3 \% & $2.8\times10^{-1}$ & $2.6\times10^{-5}$ & $4.21\times10^8$ & $2.23\times10^9$ & $5.32\times10^{10}$ & $2.98\times10^{10}$ & $1.4\times10^{-1}$ & 0.71 \\
C4$^\star$ & 60.3 \% & $1.6\times10^{-1}$ & 0.00 & $9.70\times10^8$ & $2.24\times10^9$ & $6.67\times10^{10}$ & $1.89\times10^{10}$ & $2.2\times10^{-1}$ & 0.86 \\
\hline \hline
\end{tabular}
\end{center}
$^a$ the fraction of the cumulative AGN radiant energy when $l \equiv L_{\rm BH}/L_{\rm Edd}>0.1$;\\
$^b$ the median AGN luminosity (in units of $L_{\rm Edd}$) above which AGN emits half of its total radiant energy; \\
$^c$ the median AGN luminosity (in units of $L_{\rm Edd}$) above which AGN spends half of the simulation time; \\
$^d$ black hole mass growth; \\
$^e$ total wind mass ejected by the AGN; \\
$^f$ total wind mass expelled out of the host galaxy; \\
$^g$ total star formation; \\
$^h$ the time-averaged star formation rate in the last 2 Gyr; \\
$^i$ the size of the circumnuclear disk at the end of the simulations.\\
$^\star$ the experimental run C4 stops at $t=7.3$ Gyr.
\end{table}

\textcolor{black}{
In Model C1, we turn off galaxy rotation by setting the rotation parameter $k$ 
to be zero (cf Equation \ref{eq:stellar-rotation}), i.e., it degenerates to the case
of spherical symmetry. Of course, no cold gaseous disk is formed. 
Black hole feeding is mainly via accreting cold filaments.
Significant black hole mass growth is allowed when compared to the fiducial model,
while star formation decreases by a factor of $10^5$.
No obvious correlation between black hole growth and star burst has been found in this simulation
(see also, e.g., \citealt{yuan_active_2018}).
}

\textcolor{black}{
In Model C2, we disable the effects of the Toomre instability 
(both angular momentum transfer and star formation). 
Black hole mass growth and star formation are significantly suppressed. 
The total mass of the circumnuclear disk keeps growing continuously, 
no (quasi-)steady state is found (note that there is still the $\alpha$ viscosity
in the model setup). 
The cold gaseous disk is more massive and larger than observed in normal ellipticals.
}

\textcolor{black}{
In Model C3, the $\alpha$ viscosity is turned off 
while all other physical processes are kept unchanged. 
We find fewer bursts of star formation and AGN activities in this run. 
At the late stage when the cooling flow is weak, the circumnuclear disk tends to 
sit there without radial mass transport in most of time, as  its surface density 
is not high enough to trigger the Toomre instability and thus the angular momentum transfer.
}

\textcolor{black}{
In Model C4, the hot mode feedback is disabled,
i.e., 
both the AGN luminosity and the velocity of the nuclear wind are set to be zero
when the AGN is in the hot mode. 
In this run, the low envelop of the black hole accretion rate, as a function of time, increases, 
and more black hole growth is  via ``low-level" accretion
(see also Yoon et al. 2019, in preparation).
}

\textcolor{black}{
To sum up, star formation becomes much more efficient in consuming cold gas because of galaxy rotation, when compared to the case of spherical symmetry. The Toomre instability is responsible for transferring angular momentum in the circumnuclear disk, which is crucial for both the black hole feeding and star formation. 
Though the effect of $\alpha$ viscosity are weak, it is important to notice that it could still produce indirect effects on the ``secular" evolution of the circumnuclear disk, i.e. continuously transferring angular momentum and allowing mass accretion onto the galaxy center, which may in turn induce the Toomre instability indirectly in the inner disk.  
Note that, with rotation, the fiducial model has the lowest rate of late star formation in best accord with observations
(\citealt{ford_direct_2013}).
}

\ 

\section{Discussion and Conclusion}

In this paper, we have improved our \texttt{MACER} (Massive AGN Controlled Ellipticals 
Resolved) code, and perform 2.5-dimensional simulations on the ISM 
fluid dynamics in a rotating massive elliptical galaxy. The code is 
grid-based and has high spatial resolution (parsecs in the inner regions), 
where the Bondi radius is readily resolved. The computational domain
reaches to 250 kpc, which is large enough to enclose the whole 
massive elliptical galaxy. Both passive and active stellar evolution 
are considered, and also are the mass sources from the outer and 
inner boundaries. By solving the hydrodynamics of the ISM with reasonable 
treatments of the thermal (radiative) and kinetic processes, we are able to
resolve the cooling flow directly down to (and within) the Bondi radius, therefore, 
the mass accretion rate onto the supermassive black hole is determined 
self-consistently, which is critical to evaluate the AGN feedback, 
and the latter is also included in the code. The black hole mass growth 
is tracked during the cosmological evolution of its host galaxy, 
which makes it possible to study their coevolution in a single simulation
\citep{fabian_observational_2012, kormendy_coevolution_2013}. 
Compared to our previous work (e.g. \citealt{gan_active_2014, yoon_active_2018}), 
the code has been improved comprehensively as outlined below.

We improve the galaxy modeling from a spherical configuration to allowing 
flattening and rotation \citep[see also, e.g. ][]{ciotti_effect_2017}. 
The galaxy profile is extremely important in the numerical experiments 
as it determines the characteristic temperature, velocity and timescales 
of the whole system.  In this paper, we use fully analytical axisymmetric 
models obtained by homeoidal expansion of the two-component spherical models 
(Ciotti \& Ziaee Lorzad 2019, in preparation),
which allows us to parameterize 
the galaxy morphology and its ordered rotation easily.  As the stellar winds 
inherit the velocity of their host stars, the angular momentum of the ISM 
is determined self-consistently. The rotation profile of the ISM alters 
the fluid dynamics completely by impeding the gas from being accreted, 
leading to the formation of a circumnuclear disk, and favoring star formation 
in the disk. 
\textcolor{black}{Similar behaviors can be also found in the SPH simulations by \citet{eisenreich_active_2017}, in which circumnuclear disks are commonly formed 
in the galaxy centers, and star formation occurs in those disks.}
As we have demonstrated, star formation is efficient enough to consume 
most of the cooled ISM before it could be accreted by the supermassive black hole
(see also \citealt{li_stellar_2018}). 
So, one needs to consider angular momentum transfer to study the black hole 
feeding process, and the tough competition between angular momentum transfer 
and star formation ultimately determines the fate of the gas 
in the circumnuclear disk.

We propose a numerical algorithm to compute the angular momentum transfer 
due to the classic Toomre instability. Because of the angular momentum 
barrier, the ISM will condense onto the circumnuclear disk and cool down 
further there. As a result, the disk surface density increases. 
The cold circumnuclear disk becomes gravitationally unstable 
when its surface density is higher than some critical value, 
then spiral waves will develop because of the asymmetric gravitational torque, 
which are capable of transferring angular momentum outward and making 
mass inflowing possible. The Toomre $Q$ parameter of the disk is evaluated 
instantaneously, and it is subject to the Toomre instability for those 
individual disk rings with $Q<1$. We propose that the transfer rates 
of mass and angular momentum are proportional to $\Delta Q = {\rm max}(1-Q,0)$, 
and the timescale is comparable to the local orbital time.  
As mass accretion typically occurs when the circumnuclear disk is Toomre unstable 
(with some surface density threshold), the black hole feeding is always bursty.
The cool, rotationally supported inner disk is also assumed to be MRI unstable 
and supports a weak $\alpha$ (=0.03) modulated viscosity which can transfer
angular momentum in the absence of the Toomre instability.

We improve our standard star formation algorithm (based on local cooling 
and Jeans timescales) with low-temperature and high-density thresholds 
to mimic the conditions in star forming molecular clouds. 
For the star formation in the circumnuclear disk, we also propose 
an algorithm based on the Toomre Q parameter, i.e., the gravitation 
instability drives both angular momentum transfer and star formation with 
similar timescales. As we have demonstrated, the competition between 
angular momentum transfer and star formation is critical. 
On the other hand, it is natural in our model that AGN bursts 
usually accompany strong star bursts.

We use and modify the two-mode AGN feedback model as in 
\citet{yuan_active_2018}. For the cold mode 
(high accretion rate; quasars), the implementation of AGN feedback  
is designed to match observed BAL winds and luminous output. 
For the hot mode (low accretion rate; low-luminosity AGNs), 
the properties of wind are usually hard to be measured,
so we propose the AGN feedback according to 
our knowledge gained from the theoretical studies 
(see \citealt{yuan_numerical_2015} for details).
The use of the two-mode scenario is that from both theoretical and 
observational studies we know black hole accretion has two modes 
and in each mode the descriptions of AGN outputs are very different 
(see \citealt{yuan_hot_2014} for a review).
The driving mechanisms of disk wind are also very 
different before/after the transition, especially in the hot mode, 
the wind mass loading rate is usually much larger than the black hole 
accretion rate.

We consider various mass sources including the CGM infall. 
It is important because the mass supply 
is comparable to that from stellar mass loss. With AGN/SN feedback, 
we can track the mass inflow/outflow at the galaxy outskirts. 
This also makes it possible for us to track the metal enrichment 
in/around the galaxy, which is reserved for our future \textcolor{black}{work}. 

\

With the improved code above, we investigate the cosmological evolution 
of massive elliptical galaxies in detail. We find that the results 
agree reasonably well with observations 
(e.g., \citealt{davis_atlas3d_2014,davis_wisdom_2017}):
\begin{enumerate}
\item 
	Both AGN activity and star formation are primarily in 
	central circumnuclear disks 
	(in agreement with observations, \citealt{van_dokkum_dust_1995})
	and mainly driven by the Toomre instability which are prone to be bursty, 
	and they are associated with each other. Most of the gas 
	on the disk is consumed by star formation before it can 
	be accreted by the supermassive black hole. 
\item 
	The AGN duty cycle agrees well with the Soltan argument, 
	i.e., the AGN spends most of its lifetime when it is 
	in low luminosity, while emitting most of its energy 
	when it is in high luminosity \citep{soltan_masses_1982,
	yu_observational_2002}; 
\item 
	The total star formation is $\sim$ few percents of 
	the initial stellar mass occurring in the bursts that 
	would be associated with the observed E+A phenomenon 
	\citep{dressler_spectroscopy_1982}. 
	Most of the star formation occurs in the circumnuclear 
	disk of a size $\le 1$ kpc, which is in agreement with 
	recent observations (e.g., \citealt{tadaki_gravitationally_2018}).
\item 
	The ISM X-ray luminosity varies within a reasonable range 
	and agrees well with observations.
\end{enumerate}


In our current model setup, we do not include effects of dust, 
nor any background radiation from the stars or from X-ray binaries, 
which might be worthy of consideration in the future. 
An important process that we cannot easily include is late epoch minor mergers. 
These significantly increase the mass of the high mass ellipticals (cf. \citealt{oser_two_2010} ) 
with the addition primarily of low metallicity, old stars \citep{van_dokkum_forming_2015}  
from accreted dwarf systems that puff up the outer stellar envelope 
(cf \citealt{van_dokkum_forming_2015,greene_growth_2009}), and increase the Sersic index, 
but do not greatly alter the structure within $R_e$. 
In the near future, we will perform detailed analysis on our simulation data 
and compare with observations. We will also simulate the evolution 
of the gas metallicity in/around the simulated galaxies by tracking 
the metal enrichment from the stellar/SNe winds and metal dilution 
by the CGM infall.

\section*{Acknowledgement}
We thank Ena Choi for sharing the CGM infall data. 
We thank Gregory S. Novak for sharing the first 2D version of 
the \texttt{MACER} code in 2011, which was using \texttt{ZEUSMP/1.5}.
We thank Jeremy Goodman, James Stone, Silvia Pellegrini, Pieter van Dokkum,
Nadia Zakamska, Luis Ho, Ena Choi, and Doosoo Yoon for useful discussions. 
ZG and FY are supported in part by the National Key
Research and Development Program of China (Grant
No. 2016YFA0400704), the Natural Science Founda-
tion of China (grants 11573051, 11633006, 11661161012), 
the Key Research Program of Frontier Sciences of CAS 
(No. QYZDJSSW-SYS008), the Natural Science Foundation 
of Shanghai (grant 18ZR1447200), and the Astronomical 
Big Data Joint Research Center co-founded
by the National Astronomical Observatories, 
Chinese Academy of Sciences and the Alibaba Cloud. 
This work was done during ZG's visit to the department of astronomy 
in Columbia University, which is supported by the Chinese Academy 
of Sciences via the visiting scholar program. 
We acknowledge computing resources from Columbia University's 
Shared Research Computing Facility project, which is supported by 
NIH Research Facility Improvement Grant 1G20RR030893-01, and 
associated funds from the New York State Empire State Development, 
Division of Science Technology and Innovation (NYSTAR) Contract C090171, 
both awarded April 15, 2010. Some of the simulations presented
were performed with the computing resources made available via the 
Princeton Institute for Computational Science and Engineering.

\ 

\bibliography{2018Paper1}  
\bibliographystyle{apj}

\begin{appendix}
\section{Radiative Heating/Cooling under AGN Irradiation} 
\label{appendix:radiation-processes}

In the energy equation, $H$ and $C$ are the radiative heating and cooling, 
respectively, including the contribution from AGN feedback.  
We use the formula from \citet{sazonov_radiative_2005}, 
\begin{equation}
 H - C = n^2 (S_{\rm comp} + S_{\rm brem} + S_{\rm line}),
\label{eq:radiation-heating-cooling}
\end{equation}
which includes Compton heating/cooling $S_{\rm comp}$, Bremsstrahlung cooling 
$S_{\rm brem}$, and line heating (photoionization)/cooling (recombination) 
$S_{\rm line}$ (see also \citealt{ciotti_agn_2012}). 
$n$ is the H nuclear (number) density. 
\textcolor{black}{The solar metal abundance is assumed in the calculations above.}

Here we briefly introduce the radiative processes in 
Equation \ref{eq:radiation-heating-cooling}, which includes the contributions 
from both AGN irradiation and the local atomic processes 
(we refer the readers to \citep{sazonov_radiative_2005} for details):
\begin{enumerate}
\item 
Local Bremsstrahlung cooling. 
\begin{equation} \label{eq:bremsstrahlung}
S_{\rm brem} = -3.8\times 10^{-27}\sqrt{T} \quad {\rm erg} \cdot {\rm cm}^3 /s
\end{equation}

\item
Comptonization. It could be either heating or cooling determined 
by the AGN radiation temperature $T_{\rm X}$ 
(given by Equation \ref{compton-temperature}).
\begin{equation} \label{eq:comptonization}
S_{\rm comp} = 4.1\times 10^{-35} (T_{\rm X} -T)\,\xi \quad {\rm erg} \cdot {\rm cm}^3 /s
\end{equation}

\item
Photoionization heating $S_{\rm photo}$ and recombination cooling 
$S_{recomb}$, i.e., $S_{\rm line} = S_{\rm photo} + S_{\rm recomb}$, 
where we use fitting functions below,
\begin{equation} \label{eq:line-recomb}
	S_{\rm recomb} =  10^{-23}{a \over 1 + (\xi/\xi_0)^c}   
	\quad {\rm erg} \cdot {\rm cm}^3 /s 
\end{equation}
\begin{equation} \label{eq:line-photo}
	S_{\rm photo} =  10^{-23}{ b\, (\xi/\xi_0)^c\over 1 + (\xi/\xi_0)^c}  
	\quad {\rm erg} \cdot {\rm cm}^3 /s
\end{equation}
and
\begin{equation}
	a=-{18\over  e^{25 (\log T -4.35)^2}}  -{80\over  e^{5.5(\log T -5.2)^2}} 
		-{17\over  e^{3.6(\log T -6.5)^2}},
\end{equation}
\begin{equation}
	b={1.7\times 10^4\over T^{0.7}}, 
\end{equation}
\begin{equation}
	c=1.1-{1.1\over  e^{T/1.8\,10^5}}+{4\times 10^{15}\over T^4}, 
\end{equation} 
\begin{eqnarray}
	\xi_0 &=& {1\over 1.5/\sqrt{T}+1.5\times 10^{12}/\sqrt{T^5}}+
   	     {4\times 10^{10}\over T^2} \left[1 + {80\over e^{(T-10^4)/1.5\,10^3}}\right].
\end{eqnarray}
\end{enumerate}

The effects of AGN irradiation involve equations 
\ref{eq:comptonization}-\ref{eq:line-photo} via 
the ionization parameter $\xi$, 
\begin{equation} \label{eq:ionization-parameter}
\xi \equiv \frac{L^{\rm eff}_{\rm BH, photo}(r)}{n~r^2}
\end{equation}
To evaluate the local photoionization luminosity 
$L^{\rm eff}_{\rm BH, photo}(r)$, we integrate the radial radiation 
transfer equation below \citep{ciotti_agn_2012},
\begin{equation} 
\frac{d L^{\rm eff}_{\rm BH, photo}(r)}{d r} = - 4 \pi r^2 H
\end{equation}
where the radiative heating term $H$ is ultimately determined by 
Equations \ref{eq:comptonization}-\ref{eq:line-photo}, i.e.,
\begin{equation}
H = S_{\rm photo} + {\rm max}(S_{\rm comp}, 0)
\end{equation}
Note that the formulae above are valid only when $T \geq 10^4$ K. 
Numerically, we set a temperature floor of $5\times10^3$ K for 
the self-consistency. 

\

\section{Stellar Feedback} \label{appendix:stellar-feedback}
Following \citet{ciotti_agn_2012} and \citet{pellegrini_hot_2012}, we include both the passive stellar evolution (AGBs and SNe Ia) and the 
active stellar evolution (SNe II from the newly formed stellar population). 
It is well known dying AGB stars eject winds (mass) and SNe eject huge amount 
of energy, which are recycled by the galaxy and plays an essential role 
in the galaxy evolution. The mass from AGB winds is far more than enough 
for feeding the supermassive black hole, and SNe are of capacity in 
heating up the ISM to the local Viral temperature. 
So, those processes must be considered in the galaxy evolution modeling.

As in \S\ref{sec:star-formation}, we allow star formation in our simulations. 
In the newly formed stars, we assume that a considerable faction ($20\%$) 
of the newly formed star is high-mass star ($M > 8 M_{\odot}$), 
and will turn to SN II in a timescale of $\tau_{\rm II} \simeq 2\times10^7$ 
year. We parameterize the SN II feedback as follows, 
\begin{equation} \label{eq:snii-feedback}
	\dot{\rho}_{\rm II} = \frac{\alpha_{\rm II}}{\tau_{\rm II}}\int^t_0 \dot{\rho}^
		+_\star (t^\prime) \cdot e^{-\frac{t-t^\prime}{\tau_{\rm II}}} dt^\prime, 
		\quad\quad
	\dot{E}_{\rm II} = \eta_{\rm SN} \cdot \frac{\epsilon_{\rm II} c^2}{\tau_{\rm II}}
                  \int^t_0 \dot{\rho}^+_\star (t^\prime) 
                  \cdot e^{-\frac{t-t^\prime}{\tau_{\rm II}}} dt^\prime,
\end{equation}
where $\alpha_{\rm II}$ is the ratio of SNe II mass ejecta to the total 
star formation mass,  $\epsilon_{\rm II}$ is the SN II energy efficiency.  
Following \citet{ciotti_agn_2012}, we assume (1) the newly formed stars 
is of a Salpeter IMF; (2) each massive star leaves a remnant of $1.4M_\odot$; 
(3) each SN II explosion release energy of \textcolor{black}{$10^{51}$} erg. 
We could get $\alpha_{\rm II}=0.2$ and \textcolor{black}{$\epsilon_{\rm II} = 1.9\times10^{-5}$}.

We parameterize the SNe Ia rate as
\begin{equation} \label{eq:snia-rate}
	R_{\rm SN}(t) = 0.32\times10^{-12} h^2 
		\frac{L_{\rm B}}{L_{\rm B \odot}} \left(\frac{t}{13.7 {\rm Gyr}}\right)^{-1.1} 
			\quad\quad {\rm year}^{-1}, 
\end{equation}
where $h=H_0/100~{\rm km~s^{-1}~Mpc^{-1}}$. We assume each SN Ia event 
releases $\Delta E_{\rm I} = 10^{51}$ erg of energy and ejects 
$\Delta M_{\rm I} = 1.4M_\odot$ of material into the ISM 
(i.e., the energy efficiency $\epsilon_{\rm I} 
\equiv \Delta E_{\rm I}/\Delta M_{\rm I} c^2 = 3.996\times10^{-4}$). 
Similarly, we calculate the mass and energy injection of unit volume as, 
\begin{equation} \label{eq:snia-feedback}
\dot{\rho}_{\rm I} = \Delta M_{\rm I} \frac{R_{\rm SN}}{M_\star} \rho_\star, \quad\quad
\dot{E}_{\rm I}  = \eta_{\rm SN} \cdot \epsilon_{\rm I}~\dot{\rho}_{\rm I} c^2,
\end{equation}
where $\eta_{\rm SN}$ \textcolor{black}{in the equation above (and in \ref{eq:snii-feedback}) is the SN} energy dissipation efficiency to the ISM. 
And we need the mass-to-light ratio $\Gamma \equiv \Ms/L_B$ to normalize 
Equation \ref{eq:snia-rate}. We usually set $\eta_{\rm SN} = 0.85$ 
and $\Gamma = 5.8$ in solar unit.

Following \citet{ciotti_winds_1991} we evaluate the stellar mass \textcolor{black}{loss} 
according to the stellar evolution theory, and  assume a Salpeter 
initial mass function (see also \citealt{ciotti_agn_2012}, \citealt{pellegrini_hot_2012}),
\begin{equation}\label{eq:agb-mass-loss}
\dot{M}_\star = {\rm IMF}(M_{\rm TO}) |\dot{M}_{\rm TO}| \Delta M,
\end{equation}
\textcolor{black}
{where the turn-off mass $M_{\rm TO}$ and its mass loss $\Delta M$
(in units of $M_\odot$) at time t (in units of Gyr) are, respectively,}
\begin{equation}
\textcolor{black}
{\log M_{\rm TO} = 0.0588(\log t)^2 - 0.3336 \log t + 0.2418,}
\end{equation}
\begin{equation}\label{eq:stellar-wind}
\Delta M = \cases{
	0.945 M_{\rm TO} - 0.503 ,     
		\quad\quad  M_{\rm TO}  <  9, \cr
	M_{\rm TO} - 1.4,     
		\quad\quad\quad\quad\quad\,\,  M_{\rm TO} \geq  9.    }
\end{equation}
Then, we calculate the local stellar mass loss $\dot{\rho}_\star$ 
by scaling Equation \ref{eq:agb-mass-loss} with the stellar mass density 
$\rho_\star$ (cf Equation \ref{eq:JJ2}), and evaluate the thermalization 
of the stellar mass loss according to its velocity dissipation
(Equation \ref{eq:stellar-velocity-dissipation}, see also \citealt{ciotti_effect_2017}), i.e.,
\begin{equation} \label{eq:agb-mass-loss-rate}
\dot{\rho}_\star = \rho_\star \cdot \dot{M}_\star/M_\star    \quad \quad
\dot{E}_S = {1\over2}~(\dot{\rho}_\star+\dot{\rho}_{I}+\dot{\rho}_{II}) 
		\cdot\left[ {\rm Tr}(\sigma^2) + 
		  \Vert{\bf v}-\vphi{\bf e}_{\varphi}\Vert^2\right]. 
\end{equation}
In the equation above, we assumed the stellar mass loss inherits the ordered 
rotation velocity of its host stars (cf Equation \ref{eq:stellar-rotation}). 
Finally, we inject momentum associated with the stellar mass loss accordingly,
\begin{equation} 
\dot{m}_S = (\dot{\rho}_\star+\dot{\rho}_{I}+\dot{\rho}_{II}) \cdot v_{\varphi\star}. 
\end{equation}

\end{appendix}

\end{document}